\documentclass[11pt,journal,draftcls,onecolumn,peerreviewca]{IEEEtran}
\usepackage{amsfonts}
\usepackage{amsmath}
\usepackage{amssymb}
\usepackage{bm}
\usepackage{xcolor,graphicx,float}
\usepackage{color, soul}
\usepackage{stfloats}
\usepackage[numbers,sort&compress]{natbib}
\usepackage[amsmath,thmmarks]{ntheorem}
\usepackage{theorem}
\usepackage{algorithm}
\usepackage{algorithmic}
\usepackage{graphicx}
\usepackage{subfigure}
\usepackage{pict2e}
\usepackage{comment}
\usepackage{makecell}

\theoremheaderfont{\sc}\theorembodyfont{\upshape}
\theoremstyle{nonumberplain}
\theoremseparator{}
\theoremsymbol{\rule{1ex}{1ex}}

\begin{document}
\title{Joint CFO, Gridless Channel Estimation and Data Detection for Underwater Acoustic OFDM Systems}
\author{Lei Wan, Jiang~Zhu, En Cheng and Zhiwei Xu
\thanks{Lei Wan and En Cheng are with Department of Information and Communication Engineering and Key Laboratory of Underwater Acoustic Communication and Marine Information Technology (Xiamen University), Ministry of Education, China (email: \{leiwan, chengen\}@xmu.edu.cn). Jiang Zhu and Zhiwei Xu are with the engineering research center of oceanic sensing technology and equipment, Ministry of Education, Ocean College, Zhejiang University, and also with the key laboratory of ocean observation-imaging tested of Zhejiang Province, No.1 Zheda Road, Zhoushan, 316021, China (email: \{jiangzhu16, xuzw\}@zju.edu.cn). The corresponding author is Jiang Zhu (email: jiangzhu16@zju.edu.cn). }}

\maketitle
\begin{abstract}
In this paper, we propose an iterative receiver based on gridless variational Bayesian line spectra estimation (VALSE) named JCCD-VALSE that \emph{j}ointly estimates the \emph{c}arrier frequency offset (CFO), the \emph{c}hannel with high resolution and carries out \emph{d}ata decoding. Based on a modularized point of view and motivated by the high resolution and low complexity gridless VALSE algorithm, three modules named the VALSE module, the minimum mean squared error (MMSE) module and the decoder module are built. Soft information is exchanged between the modules to progressively improve the channel estimation and data decoding accuracy. Since the delays of multipaths of the channel are treated as continuous parameters, instead of on a grid, the leakage effect is avoided. Besides, the proposed approach is a more complete Bayesian approach as all the nuisance parameters such as the noise variance, the parameters of the prior distribution of the channel, the number of paths are automatically estimated. Numerical simulations and sea test data are utilized to demonstrate that the proposed approach performs significantly better than the existing grid-based generalized approximate message passing (GAMP) based \emph{j}oint \emph{c}hannel and \emph{d}ata decoding approach (JCD-GAMP). Furthermore, it is also verified that joint processing including CFO estimation provides performance gain.
\end{abstract}
{\bf keywords:} Iterative receivers, message passing, expectation propagation, grid-less, OFDM channel estimation
\section{Introduction}

\IEEEPARstart{D}{ue} to the propagation characteristics, underwater acoustic communications typically suffer from strong multipath and Doppler effect~\cite{StPr09}. Since the propagation speed is low (around 1500 m/s), the reflection of acoustic waves from surface and bottom of the ocean, along with the refraction caused by nonuniformity of sound speed in water, cause significant multipath delay spread. The typical delay spread of underwater acoustic communication channels is from several milliseconds to tens of milliseconds, and could be hundreds of milliseconds and more in extreme cases. On the other hand, due to the movement of transmitter and receiver platforms, as well as water media, Doppler effect is almost inevitable in underwater acoustic communications. Considering the wide band nature of underwater acoustic communication systems, the Doppler effect usually exhibits itself as signal dilation or compression~\cite{Sharif00JOE}.

With the advantages of low complexity frequency domain multipath channel equalization and high bandwidth efficiency, orthogonal frequency division multiplexing (OFDM) technology was introduced into underwater acoustic communications, and soon became a popular choice especially for high speed communications~\cite{Stoj06,Zhou08JOE,zhya20}. However, Doppler effect in underwater acoustic channels could destroy the orthogonality between the subcarriers in OFDM and hence deteriorates its performance. In~\cite{Zhou08JOE}, a two-step approach was proposed to compensate the Doppler distortion, in which a resampling operation is carried out first to convert the wide band Doppler distortion into narrow band carrier frequency offset (CFO). In the second step, the CFO is estimated and compensated. Experimental results in~\cite{Zhou08JOE} demonstrated that the proposed scheme can effectively mitigate the Doppler distortion.

In coherent underwater acoustic communication systems, accurate channel estimation is essential. Since compressed sensing based channel estimation methods exploit the structure of the channel and require less number of measurements, they have been the most popular choices for underwater acoustic OFDM systems~\cite{LiPr07, bzpw10}. Among compressed sensing algorithms, orthogonal matching pursuit (OMP) algorithm~\cite{trgi07} features lower computational complexity than basis pursuit methods, and it is widely adopted in channel estimations~\cite{prdk15, paup19}. On the other hand, as an approximation for sum-product algorithm (SPA), approximate message passing (AMP) has been utilized for sparse linear estimation problems~\cite{AMP}, and hence is also employed for sparse channel estimations~\cite{wlzd17}.

To further improve the performance of wireless communication systems in challenging channel conditions, joint receivers in which modules such as channel estimation, channel equalization interactively work together are proposed, both for radio and underwater acoustic communication systems. In~\cite{pasp10}, a joint channel estimation, equalization and data detection system based on expectation-maximization (EM) algorithm is proposed for OFDM systems with high mobility. In the proposed scheme, the time-varying channel is represented by discrete cosine orthogonal basis functions and the equalized symbols are quantized into nearest data symbol constellation point in each iteration. A low complexity generalized AMP (GAMP) based joint channel estimation and data decoding approach for OFDM system is proposed \cite{SchniterJAMP}. Later, a joint channel estimation and symbol detection scheme based on parametric bilinear GAMP algorithm is proposed for single-carrier systems \cite{Schniter_AMP18}, which supports the sparse characteristic of wireless communication channels and soft-input soft-output (SISO) equalizer. Recently, based on Markov chain Monte Carlo (MC) algorithm with Gibbs sampling and convolutional coding plus differential coding, a joint phase shift keying (PSK) data detection and channel estimation scheme utilizing channel sparsity is proposed in~\cite{jswd20}.

For underwater acoustic communications, the concept of joint processing has been widely pursued. In the early work~\cite{sshk08}, an iterative receiver based on message passing techniques is proposed for single carrier underwater acoustic communication systems, in which equalization and decoding are performed multiple times in a turbo manner. However, channel estimation is carried out based on training symbols, utilizing maximum-likelihood approach. In~\cite{yazh16,chwz17,qiqz21}, iterative receivers were developed for single carrier multiple-input multiple-output (MIMO) underwater acoustic communication systems, for both time-domain and frequency domain equalization. In the proposed iterative receivers, soft equalization outputs are feedback for interference cancelation and channel estimation. For OFDM underwater acoustic communication systems in doubly spread channels, an iterative sparse channel estimation and data detection receiver using partial interval demodulation is proposed in~\cite{armu18}. In~\cite{tuxs21}, an iterative receiver is proposed for inter-frequency interference (IFI) cancelation for single-carrier underwater acoustic communication systems adopting frequency domain equalization. Channel impulse responses, explicit and residual IFI are updated in an iterative fashion based on the soft symbol estimates in the proposed receiver. Considering the impulse noise in underwater acoustic communication channels, a joint channel estimation and impulsive noise mitigation receiver utilizing sparse Bayesian learning (SBL) algorithm is proposed \cite{Rong20}, in which the information on data symbols are employed to carry out joint time-varying channel estimation and data detection, to further improve the system performance. The performance of the proposed algorithm is verified through both numerical simulations and real data. More recently, a joint channel estimation and equalization scheme with superimposed training is proposed for single carrier underwater acoustic communication systems~\cite{ygdy21}. A message-passing-based bidirectional channel estimation algorithm is adopted in the proposed scheme, utilizing the correlation between consecutive channels.

It is worth noting that all the above compressed sensing based approaches utilize the fixed dictionary and suffer grid mismatch \cite{MMis}, and the gridless parametric channel estimation approach indeed improves both the channel estimation and data decoding performance over the grid based approaches \cite{Supfast, JointSupfast}. Moreover, existing iterative receivers seldom incorporate the CFO estimation and compensation into the design.
In this paper, according to message passing and from the modularized point of view \cite{Mengunified}, an iterative receiver based on gridless variational Bayesian line spectra
estimation (VALSE)~\cite{Badiu} which \emph{j}ointly estimates the residual {\emph C}FO and {\emph c}hannel, meanwhile also achieves {\emph d}ata decoding named JCCD-VALSE is proposed for underwater acoustic OFDM systems. The JCCD-VALSE consists of three modules named as high resolution gridless channel estimation, minimum mean squared error (MMSE) estimation, and low-density parity-check code (LDPC) decoder. Through the exchange of soft information between modules, the channel estimation and data decoding accuracy progressively improves. Compared with existing receivers, the joint processing including residual CFO estimation and compensation, the gridless channel estimation, and the feature of automatical estimation of nuisance parameters are the advantages of the proposed scheme. Hence, the contributions of this paper include:
\begin{enumerate}
  \item
  Targeting at underwater acoustic OFDM systems in channels with an approximately common Doppler scaling factor, the system model for received signal after resampling is derived, and the corresponding receiver JCCD-VALSE which jointly estimates residual CFO, channel, and finishes data decoding is proposed.
  \item
  In JCCD-VALSE, all the nuisance parameters such as the number of paths, the noise variance are automatically estimated. This reduces the reliance on priori information, and also improves the estimation accuracy.
  \item
  In JCCD-VALSE, gridless channel estimation based on VALSE is adopted, which improves the channel estimation and data decoding performance, compared with the existing channel estimation and data decoding methods. Besides, it is also shown that taking residual CFO into consideration benefits both the channel estimation and data decoding performances.
  \item
  Sea trial data of mobile acoustic communication experiment (MACE10) is utilized to further demonstrate the effectiveness of the proposed scheme.
\end{enumerate}
In addition, it is also worth noting that our proposed approach is very flexible and can deal with both uniform and nonuniform pilots.

The rest of this paper is organized as follow: Section \ref{systmodel} introduces the system model. The probabilistic formulation of the channel, the data symbol and the normalized residual CFO are presented in Section \ref{Probform}. The ensuing Section \ref{Algorithm} proposes the joint CFO, channel estimation and data decoding algorithm. Section \ref{NSres} presents the performance metrics in terms of the coded bit error rate (BER) and the normalized mean squared error (NMSE) of the various algorithms. The sea trial data decoding results are introduced in Section \ref{Realdata}. Finally, we conclude the paper in Section \ref{Con}.

Notation:
For a matrix $\mathbf A$, let $[{\mathbf A}]_{i,j}$ or $A_{ij}$ denote the $(i,j)$th element of ${\mathbf A}$, and ${\rm diag}(\mathbf A)$ returns a vector with elements being the diagonal elements of $\mathbf A$. Let $\Re\{\cdot\}$ and $\Im\{\cdot\}$ denote the real and imaginary part operator, respectively. For random vectors $\mathbf x$ and $\mathbf y$ with joint probability density function (PDF) $p({\mathbf x},{\mathbf y})$, the marginal PDF $p({\mathbf x})=\int p({\mathbf x},{\mathbf y}) {\rm d}{\mathbf y}$ with ${\rm d}{\mathbf y}$ being ${\rm d}\Re\{\mathbf y\}{\rm d}\Im\{\mathbf y\}$. Let $\odot$ denote the Hadamard product operator. For a vector ${\mathbf a}$, let ${\rm diag}({\mathbf a})$ return a diagonal matrix with diagonal elements being $\mathbf a$. Let ${\mathcal S}\subset \{1,\cdots,N\}$ be a subset of indices. For a square matrix ${\mathbf A}\in {\mathbb C}^{N\times N}$, let $\mathbf A_{{\mathcal S},{\mathcal S}}$ denote the submatrix by choosing both the rows and columns of $\mathbf A$ indexed by $\mathcal S$. Let ${(\cdot)}^{*}_{\mathcal S}$, ${(\cdot)}^{\rm T}_{\mathcal S}$ and ${(\cdot)}^{\rm H}_{\mathcal S}$ be the conjugate, transpose and Hermitian transpose operator of ${(\cdot)}_{\mathcal S}$, respectively. For two vectors ${\mathbf a}\in {\mathbb C}^n$ and ${\mathbf b}\in {\mathbb C}^n$, ${\mathbf a}/{\mathbf b}$ denotes the componentwise division.

\section{System model}\label{systmodel}
This section establishes the OFDM model similar to \cite{Zhou08JOE}, and the assumption that all paths have a similar Doppler scaling factor is adopted, which seems to be justified as long as the dominant Doppler effect is caused by the direct transmitter/receiver motion. For completeness, the detailed derivations are presented.

Consider the cyclic prefix (CP) OFDM system, and let $T$ and $T_{\rm cp}$ denote the OFDM symbol duration and the CP duration. The number of subcarriers is $N$ and the frequency spacing is $\Delta f=1/T$. With carrier frequency $f_c$, the frequency of the $k$th subcarrier is
\begin{align}
f_k = f_c + k\Delta f,~k=-N/2,\dots,N/2-1,\notag
\end{align}
and the bandwidth is $B=N\Delta f$.

Let $d[k]$ denote the information symbol to be transmitted on the $k$th subcarrier. The transmitted continuous time OFDM block in passband is
\begin{align}
s(t) = \frac{2}{\sqrt{N}}\Re \left\{ \sum\limits_{k=-N/2}^{N/2-1} d[k] {\rm e}^{{\rm j}2\pi f_k t}g(t) \right\},\notag
\end{align}
where $g(t)$ is
\begin{align}
g(t)=\begin{cases}
1, \quad t\in[-T_{\rm cp}, T],\\
0,\quad {\rm otherwise}.
\end{cases}\notag
\end{align}

A multipath underwater channel where all paths have a similar Doppler scaling factor $a$ is considered, and the channel impulse response is
\begin{align}
c(\tau,t)=\sum\limits_{p=1}^{L}A_p\delta(\tau-\tau_p(t))=\sum\limits_{p=1}^{L}A_p\delta(\tau-(\tau_p-at)),\notag
\end{align}
where $A_p$ and $\tau_p$ denote the gain and delay of the $p$th path, $L$ is the total number of paths.
The received continuous time signal in passband $\tilde{y}(t)$ is
\begin{align}
\tilde{y}(t)=s(t)\ast c(\tau,t)&= \sum\limits_{p=1}^{L}A_ps(t-\tau_p+at) \notag \\
&=\frac{2}{\sqrt{N}}\Re \left\{ \sum\limits_{k=-N/2}^{N/2-1} d[k] \sum\limits_{p=1}^{L}A_p{\rm e}^{{\rm j}2\pi f_k (t-(\tau_p-at))}g((1+a)t-\tau_p) \right\}. \notag
\end{align}
After downconverting and low pass filtering, the baseband signal ${y}(t)$ can be obtained as
\begin{align}
{y}(t)&={\rm LPF}\left[\tilde{ y}(t){\rm e}^{-{\rm j}2\pi f_ct}\right]\notag\\
&\approx\frac{1}{\sqrt{K}}\sum\limits_{k=-N/2}^{N/2-1} d[k] {\rm e}^{{\rm j}2\pi k\Delta ft}{\rm e}^{{\rm j}2\pi af_kt}\left[\sum\limits_{p=1}^{L}A_p{\rm e}^{-{\rm j}2\pi f_k \tau_p}g((1+a)t-\tau_p)\right]+n(t),\notag
\end{align}
where $n(t)$ is the additive noise in baseband. Note that each subcarrier experiences a Doppler-induced frequency shift ${\rm e}^{{\rm j}2\pi af_kt}$. Provided that the bandwidth of the OFDM signal is comparable to the center frequency, the Doppler-induced frequency shift differs considerably on different subcarriers, which introduces the intercarrier interference.

In order to mitigate the Doppler effect and convert the wideband OFDM system into a narrowband system, a resampling approach is carried out first \cite{Zhou08JOE}. Suppose the Doppler factor $a$ is estimated as $\hat{a}$.
Resample ${y}(t){\rm e}^{-{\rm j}2\pi \hat{a}f_ct}$ with a resampling factor $\hat{a}$ to yield
\begin{align}\label{origeq}
{z'}(t)&={y}(\frac{t}{1+\hat{a}}){\rm e}^{-{\rm j}2\pi \hat{a}f_c\frac{t}{1+\hat{a}}}\notag\\
&=\frac{1}{\sqrt{N}}\sum\limits_{k=-N/2}^{N/2-1} d[k] {\rm e}^{{\rm j}2\pi \frac{1+a}{1+\hat{a}}k\Delta ft}{\rm e}^{{\rm j}2\pi \frac{a-\hat{a}}{1+\hat{a}}f_ct}\left[\sum\limits_{p=1}^{L}A_p{\rm e}^{-{\rm j}2\pi f_k \tau_p}g(\frac{1+a}{1+\hat{a}}t-\tau_p)\right]+n(t){\rm e}^{{\rm j}2\pi \frac{\hat{a}}{1+\hat{a}}f_ct},
\end{align}
The resampling factor $\hat{a}$ is chosen to make  $\frac{1+a}{1+\hat{a}}$ as close to one as possible. As a result, (\ref{origeq}) can be approximated as
\begin{align}\label{ofdm_resample}
{z'}(t)\approx \frac{1}{\sqrt{N}}{\rm e}^{{\rm j}2\pi \frac{a-\hat{a}}{1+\hat{a}}f_c t}\sum\limits_{k=-N/2}^{N/2-1} d[k] {\rm e}^{{\rm j}2\pi k\Delta ft}\left[\sum\limits_{p=1}^{L}A_p{\rm e}^{-{\rm j}2\pi f_k \tau_p}g\left(t-\tau_p\right)\right]+v(t).
\end{align}
From~\eqref{ofdm_resample}, usually CFO estimation and compensation are carried out as the second step to further remove the impact of Doppler~\cite{Zhou08JOE}. Let ${r}(t)$ denote the output of the CFO compensation, sampling $r(t)$ at $(n-1)T_s=(n-1)/B$, $n=1,\dots,N$ yields \footnote{In engineering application, fftshift operator ${\rm fftshift}(\cdot)$ is usually introduced and model (\ref{compfinal}) can also be formulated as ${\mathbf r}={\mathbf e}(\omega)\odot \left({\mathbf F}^{\rm H}{\rm fftshift}({\mathbf h}\odot {\mathbf d})\right)+{\mathbf w}$.}
\begin{align}\label{compfinal}
{\rm e}^{{\rm j}\pi (n-1)}{\mathbf r}={\mathbf e}(\omega)\odot \left({\mathbf F}^{\rm H}({\mathbf h}\odot {\mathbf d})\right)+{\mathbf w},
\end{align}
where ${r}_{n}=r(t)|_{t=(n-1)T_s}$; the $n$th entry $d_n$ of ${\mathbf d}$ is $d_n=d[n-1-N/2]$; ${\mathbf F}$ denotes the normalized DFT matrix with the $(m,n)$th entry being $\frac{1}{\sqrt{N}}{\rm e}^{-{\rm j}2\pi(n-1)(m-1)/N}$; $\omega$
denotes the normalized residual CFO which is usually very small and close to $0$, ${\mathbf e}(\omega)=[1,{\rm e}^{{\rm j}\omega},\cdots,{\rm e}^{{\rm j}(N-1)\omega}]^{\rm T}$; $\mathbf h$ is the frequency-domain channel response vector and can be viewed as a line spectra, which can be represented as
\begin{align}
{\mathbf h}={\mathbf A}({\boldsymbol \theta}){\boldsymbol \beta},\label{hdef}
\end{align}
where ${{\boldsymbol \beta}\in {\mathbb C}^L}$ denotes the complex coefficient vector related to the path gains and delays of the channel and
\begin{align}
\beta_p=A_p{\rm e}^{-{\rm j}2\pi f_{-\frac{N}{2}}\tau_p},\notag
\end{align}
${{\boldsymbol \theta}\in {\mathbb R}^L}$ denotes the frequency component related to the path time delay of the channel and
\begin{align}
\theta_p=-2\pi\Delta f\tau_p,\notag
\end{align}
$L$ denotes the number of paths, ${\mathbf A}({\boldsymbol \theta})$ is
\begin{align}
{\mathbf A}({\boldsymbol \theta})=[{\mathbf a}(\theta_1),{\mathbf a}(\theta_2),\cdots,{\mathbf a}(\theta_{L})],\notag
\end{align}
and ${\mathbf a}(\theta)=[1,{\rm e}^{{\rm j}\theta},\cdots,{\rm e}^{{\rm j}(N-1)\theta}]^{\rm T}$; $\mathbf w$ denotes the noise and approximation error which is assumed to follow the Gaussian distribution, i.e., ${\mathbf w}\sim {\mathcal {CN}}({\mathbf 0},\sigma^2{\mathbf I}_N)$ with $\sigma^2$ being the unknown variance.

Define
\begin{align}
{y}_n={\rm e}^{{\rm j}\pi (n-1)}{z}_n.\notag
\end{align}
Then, (\ref{compfinal}) is simplified as
\begin{align}\label{simpfinal}
{\mathbf y}={\mathbf e}(\omega)\odot \left({\mathbf F}^{\rm H}({\mathbf h}\odot {\mathbf d})\right)+{\mathbf w}.
\end{align}


Assume that the symbol vector $\mathbf d$ are partitioned into three parts corresponding to the pilot, null, and data, which are represented as ${\mathbf d}_{{\mathcal P}}$, ${\mathbf d}_{{\mathcal N}}$ and ${\mathbf d}_{{\mathcal D}}$ and ${\mathcal P}\cup{\mathcal N}\cup{\mathcal D}=\{1,2,\cdots,N\}$. Then the goal is to jointly estimate the data symbol ${\mathbf d}_{{\mathcal D}}$, the channel vector $\mathbf h$, the normalized residual CFO $\omega$, and the nuisance parameters such as $\sigma^2$ by fully exploiting the channel structure $\mathbf h$. In the following, we design an approximate Bayesian algorithm.

\section{Probabilistic Formulation}\label{Probform}
This section describes the probabilistic formulation of the channel $\mathbf h$, the data symbol $\mathbf d$ and the normalized residual CFO $\omega$.

\subsection{Parametric Channel Model}
For the channel $\mathbf h$, the number of path $L$ is usually unknown. Here the number of paths is assumed to be $L_{\rm max}$ and $L<L_{\rm max}$ \cite{Badiu}, i.e.,
\begin{align}\label{signal-model}
{\mathbf h}=\sum\limits_{l=1}^{L_{\rm max}} {\beta}_l {\mathbf a}({\theta}_l)\triangleq {\mathbf A}({\boldsymbol \theta}){\boldsymbol \beta},
\end{align}
where ${\mathbf A}({\boldsymbol \theta})=[{\mathbf a}({\theta}_1),\cdots,{\mathbf a}({\theta}_{L_{\rm max}})]$. Since the number of path is supposed to be $L_{\rm max}$, the binary hidden variables ${\mathbf s}=[s_1,...,s_{L_{\rm max}}]^{\rm T}$ are introduced, where $s_l=1$ means that the $l$th path is active, otherwise deactive ($\beta_l=0$). Let $\rho$ denote the prior probability of each path being active, i.e.,
\begin{align}\label{sprob}
p(s_l) = \rho^{s_l}(1-\rho)^{(1-s_l)},\quad s_l\in\{0,1\}.
\end{align}
Then the probability mass function (PMF) of ${\mathbf s}$ is $p({\mathbf s})=\prod\limits_{l=1}^{L_{\rm max}}p(s_l)$. Given that $s_l=1$, we assume that ${\beta}_l\sim {\mathcal {CN}}({\beta}_l;0,\nu)$, where $\nu$ denotes the priori variance of the path being active. Thus $(s_l,{\beta}_l)$ follows a Bernoulli-Gaussian distribution, that is
\begin{align}
p({\beta}_l|s_l;\nu) = (1 - s_l){\delta}({\beta}_l) + s_l{\mathcal {CN}}({\beta}_l;0,\nu).\label{pdfw}
\end{align}
In addition, the distribution of $\boldsymbol \beta$ conditioned on $\mathbf s$ is $p({\boldsymbol \beta}|{\mathbf s})=\prod\limits_{l=1}^{L_{\rm max}}p({\beta}_l|s_l;\nu)$.

From (\ref{sprob}) and (\ref{pdfw}), it can be seen that the parameter $\rho$ denotes the probability of the $l$th component being active and $\nu$ is a variance parameter. The variable ${\boldsymbol \theta} = [\theta_1,...,\theta_{L_{\rm max}}]^{\rm T}$ has the prior PDF $p({\boldsymbol \theta}) = \begin{matrix} \prod_{l=1}^{L_{\rm max}} p(\theta_l) \end{matrix}$. Generally, $p(\theta_l)$ is encoded through the von Mises distribution \cite[p.~36]{Direc}
\begin{small}
\begin{align}\label{prior_theta}
p(\theta_l) = {\mathcal {VM}}(\theta_l;\mu_{0,l},\kappa_{0,l})= \frac{1}{2\pi{I_0}(\kappa_{0,l})}{\rm e}^{\kappa_{0,l}{\cos(\theta-\mu_{0,l})}},
\end{align}
\end{small}
where $\mu_{0,l}$ and $\kappa_{0,l}$ are the mean direction and concentration parameters of the prior of the $l$th frequency $\theta_l$, $I_p(\cdot)$ is the modified Bessel function of the first kind and the order $p$ \cite[p.~348]{Direc}. For a given von Mises distribution,
${\rm arg}({\rm E}_{{\mathcal {VM}}(\theta;\mu,\kappa)}[{\rm e}^{{\rm j}\theta}]) = {\rm arg}\left({\rm e}^{{\rm j}\mu}\frac{I_1(\kappa)}{I_0(\kappa)}\right) = \mu = {\rm E}_{{{\mathcal {VM}}(\theta;\mu,\kappa)}}[\theta]$. In addition, ${\rm E}[{\rm e}^{{\rm j}m\theta}]={\rm e}^{{\rm j}m\mu}(I_m(\kappa)/I_0(\kappa))$ \cite[pp.~26]{Direc}. Without any knowledge of the frequency $\theta_l$, the uninformative prior distribution $p(\theta_l) = {1}/({2\pi})$ is used. For more details please refer to \cite{Badiu}.

\subsection{OFDM System}
A single-input single-output OFDM system with $N$ subcarriers is considered. Since the structure of the consecutive OFDM symbols is not exploited, a single OFDM system is modeled \footnote{As you see, the proposed approach outputs the PDFs of the amplitudes and delays of all the paths, which is very suitable for channel tracking and will be left for future work.}. Of the $N$ subcarriers, $N_{\mathcal P}$ and $N_{\mathcal N}$ are dedicated as pilots and nulls, respectively, and the remaining $N_{\mathcal D}=N-N_{\mathcal P}-N_{\mathcal N}$ are used to transmit the coded/interleaved data bits. The sets $\mathcal N$, $\mathcal P$, $\mathcal D$ denote the indices of the null, pilot and data subcarriers, respectively. It follows that ${\mathcal P}\cup{\mathcal N}\cup{\mathcal D}=\{1,2,\cdots,N\}$ and ${\mathcal P}\cap{\mathcal N}=\emptyset$, ${\mathcal P}\cap{\mathcal D}=\emptyset$ and ${\mathcal N}\cap{\mathcal D}=\emptyset$. Besides, the null subcarriers are not necessary to delicate and can be set as the empty set ${\mathcal N}=\emptyset$ to improve the data transmission rate.

Let the vector ${\mathbf u}\in \{0,1\}^K$ be the transmitted (equi-probable) information bits. These bits are coded by a rate-$R$ encoder and interleaved to get the length-$K/R$ vector $c={\mathcal C}({\mathbf u})$, where ${\mathcal C}:\{0,1\}^K\rightarrow \{0,1\}^{K/R}$ is the interleaving and coding function such as a turbo code and a LDPC code. The coded/interleaved bits are partioned into subvectors ${\mathbf c}^{(i)}\in \{0,1\}^Q$, $i\in {\mathcal D}$, which are then mapped to the $i$th subcarrier. Let ${\mathcal M}:\{0,1\}^Q\rightarrow {\mathbb A}_D \subset {\mathbb C}$ denote the $2^Q$-ary mapping, where ${\mathbb A}_D$ is the symbol alphabet. The pilots are drawn from the pilot symbol alphabet ${\mathbb A}_P \subset {\mathbb C}$. In OFDM, ${\mathbb A}_D$ is typically a $2^Q$-ary quadrature amplitude modulation (QAM) alphabet and ${\mathbb A}_P$ is a quadrature phase shift keying (QPSK) alphabet. The nulls, pilots and data symbols are stacked in vector $\mathbf d$. Vectors ${\mathbf d}_{\mathcal N}$, ${\mathbf d}_{\mathcal P}$ and ${\mathbf d}_{\mathcal D}$ contain the nulls, pilot symbol and data symbols, respectively.

For the OFDM symbol, the information bits ${\mathbf u}$ has the following PDF
\begin{align}
    p({\mathbf u})=\prod\limits_{k\in {\mathcal K}}p(u_k)=\prod\limits_{k\in {\mathcal K}}\left(0.5\times {1}_{[u_k\in \{0,1\}]}\right),\notag
\end{align}
where ${1}_{[\cdot]}$ denotes the indicator function, ${\mathcal K}$ denotes the index set of the information bits. For the interleaving and coding function, the PMFs $p({\mathbf c}|{\mathbf u})$ of subvectors ${\mathbf c}$ conditioned on $\mathbf u$ are
\begin{align}
    p({\mathbf c}|{\mathbf u})={ 1}_{[{\mathbf c}=C({\mathbf u})]}.\notag
\end{align}
Note that $p({\mathbf c}|{\mathbf u})$ describe the structure of the channel code and interleaver, whose details are not provided. Conditioned on $\mathbf c$, the PMF of $\mathbf d$ is $p({\mathbf d}|{\mathbf c})$.

The normalized residual CFO $\omega$ is treated as an unknown deterministic parameter and $\omega\in [-\pi,\pi]$.

\begin{figure}[h!t]
\centering
\includegraphics[width=4.8in]{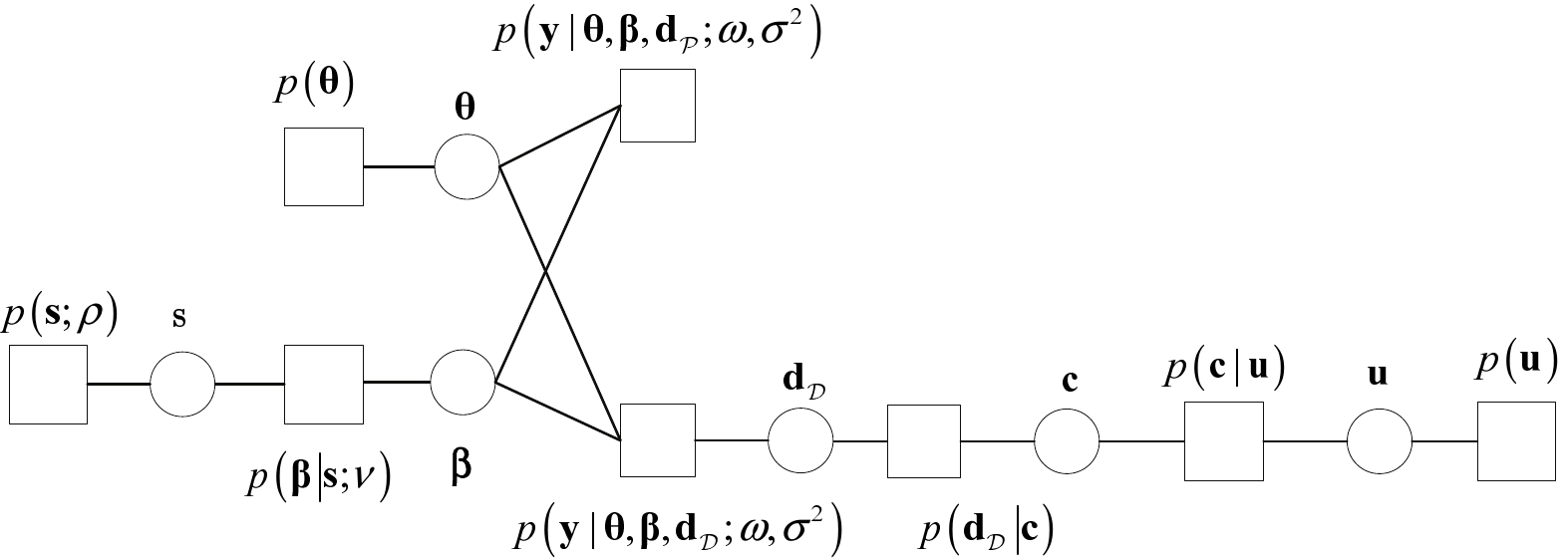}
\caption{Factor graph of the joint PDF (\ref{jointpdf}) describing the parametric channel and OFDM system. Here the circle denotes the variable (vector) node, and the square denotes the factor node. The nulls are not plotted here for simplicity, as it can only be used for noise variance estimation.}
\label{FacOrig}
\end{figure}

According to the above probabilistic model, the factor graph representation describing the channel and OFDM system model is given in Fig. \ref{FacOrig}. The joint PDF is
\begin{align}\label{jointpdf}
p({\mathbf y},{\mathbf d}_{{\mathcal D}},{\mathbf c},{\mathbf u},{\boldsymbol \beta},{\boldsymbol \theta},{\mathbf s},;\omega,\nu,\rho,\sigma^2)= p({\mathbf s};\rho)p({\boldsymbol \beta}|{\mathbf s};\nu)p({\boldsymbol \theta})
p({\mathbf u})p({\mathbf c}|{\mathbf u})p({\mathbf d}|{\mathbf c})p({\mathbf y}|{\boldsymbol \theta}, {\boldsymbol \beta},{\mathbf d}_{{\mathcal D}};\omega,\sigma^2).
\end{align}
Firstly, the maximum likelihood estimation of the nuisance parameters are
\begin{align}\label{MLnupa}
(\omega_{\rm {ML}},\nu_{\rm {ML}},\rho_{\rm {ML}},\sigma_{\rm {ML}}^2)&=\underset{\omega,\nu,\rho,\sigma^2} {\operatorname {argmax}}\quad p({\mathbf y};\omega,\nu,\rho,\sigma^2)\notag\\
&=\int p({\mathbf y},{\mathbf d}_{{\mathcal D}},{\mathbf c},{\mathbf u},{\boldsymbol \beta},{\boldsymbol \theta},{\mathbf s},;\omega,\nu,\rho,\sigma^2){\rm d}{\mathbf d}_{{\mathcal D}}{\rm d}{\boldsymbol \beta}{\rm d}{\mathbf u}{\rm d}{\mathbf c}{\rm d}{\boldsymbol \theta}{\rm d}{\mathbf s}.
\end{align}
Then the BER optimal receiver, i.e., maximum a posterior (MAP) estimate is
\begin{align}\label{MAP_rec}
\hat{u}_k=\underset{u_k\in\{0,1\}}{\operatorname{argmax}}~p(u_k|{\mathbf y};\omega_{\rm {ML}},\nu_{\rm {ML}},\rho_{\rm {ML}},\sigma_{\rm {ML}}^2),\end{align}
where $p(u_k|{\mathbf y};\omega_{\rm {ML}},\nu_{\rm {ML}},\rho_{\rm {ML}},\sigma_{\rm {ML}}^2)\propto p(u_k,{\mathbf y};\omega_{\rm {ML}},\nu_{\rm {ML}},\rho_{\rm {ML}},\sigma_{\rm {ML}}^2)$ can be obtained via marginalizing all the random variables but $u_k$ in the joint PDF $p({\mathbf y},{\mathbf d}_{{\mathcal D}},{\mathbf c},{\mathbf u},{\boldsymbol \beta},{\boldsymbol \theta},{\mathbf s},;\omega_{\rm {ML}},\nu_{\rm {ML}},\rho_{\rm {ML}},\sigma_{\rm {ML}}^2)$. Note that solving either (\ref{MLnupa}) or (\ref{MAP_rec}) is computationally intractable. Therefore, we resort to approximate Bayesian methods in the ensuing section.

\section{Algorithm}\label{Algorithm}
This section utilizes expectation propagation (EP) \cite{Minka} to develop the JCCD-VALSE algorithm through novelly combining the gridless VALSE, the MMSE module and the LDPC decoder module.

\begin{figure}[h!t]
\centering
\includegraphics[width=5.2in]{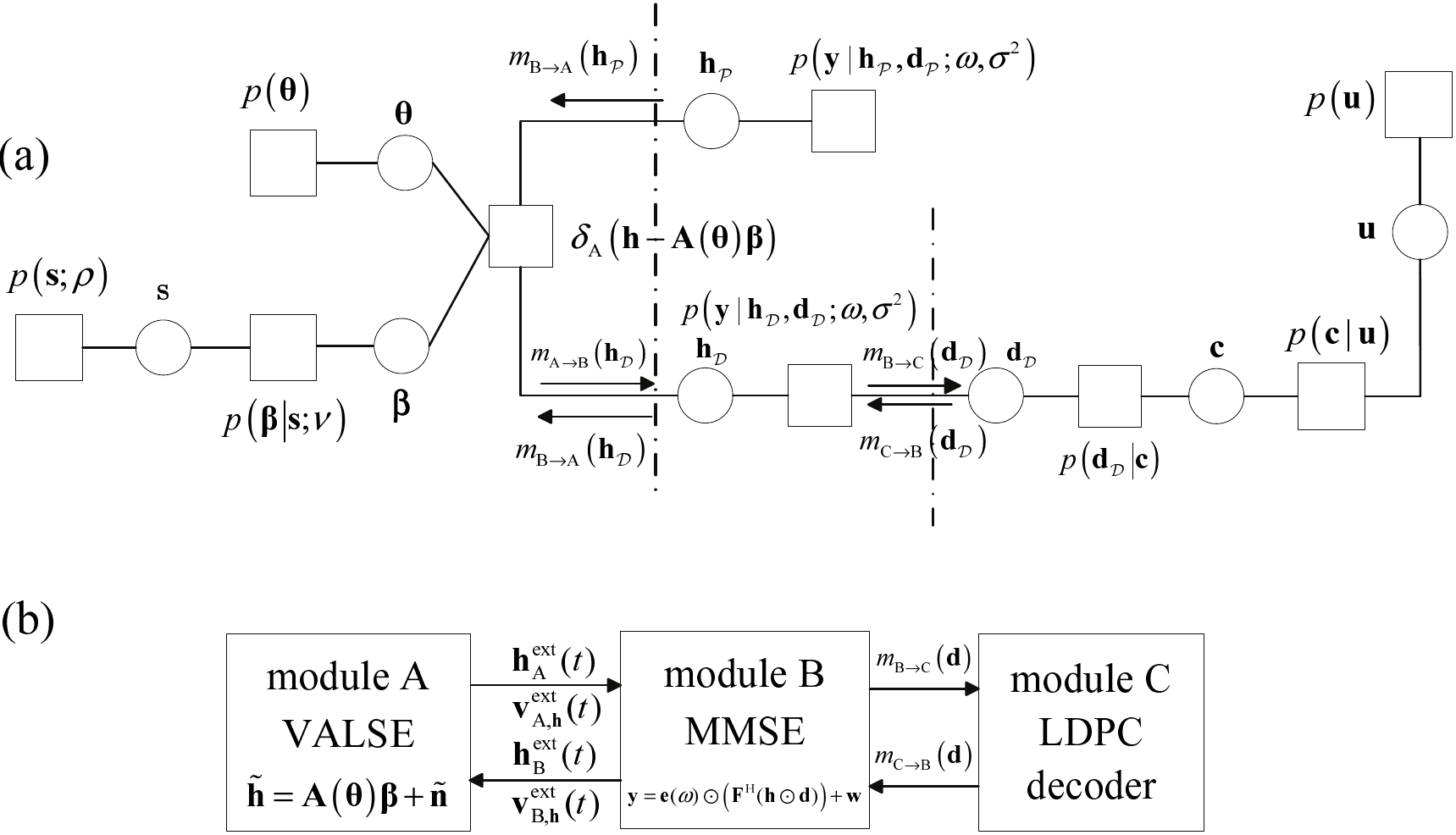}
\caption{Factor graph of the joint PDF (\ref{jointpdf}) with an additional hidden variable node $\mathbf h$ and the module of the JCCD-VALSE algorithm. This factor graph is equivalent to the factor graph shown in Fig. \ref{FacOrig}. }
\label{FC_turbo_fig}                        \end{figure}

It is worth noting that in conventional underwater OFDM systems, null subcarriers are beneficial for both Doppler and noise estimation. While in this work, the pilot and data can be used for both Doppler and noise estimation.  As a consequence, the null subcarriers can be replaced with the data subcarriers, which improves the spectrum efficiency. Define the ordered set ${\mathcal M}={\mathcal P}\cup {\mathcal D}$.

Before designing the algorithm, an important point is emphasized. Note that
\begin{align}
p({\mathbf y}|{\mathbf h},{\mathbf d};\omega,\sigma^2)&=\frac{1}{(\pi\sigma^2)^N}{\rm e}^{-\frac{\left({\mathbf y}-{\mathbf e}(\omega)\odot \left({\mathbf F}^{\rm H}({\mathbf h}\odot {\mathbf d})\right)\right)^{\rm H}\left({\mathbf y}-{\mathbf e}(\omega)\odot \left({\mathbf F}^{\rm H}({\mathbf h}\odot {\mathbf d})\right)\right)}{\sigma^2}}\notag\\
&=\frac{1}{(\pi\sigma^2)^N}{\rm e}^{-\frac{\|\tilde{\mathbf y}(\omega)-{\mathbf h}\odot {\mathbf d}\|_2^2}{\sigma^2}}\notag\\
&=\prod\limits_{i=1}^Np(\tilde{y}_i(\omega)|{h}_i,{d}_i;\omega,\sigma^2),
\end{align}
where $\tilde{\mathbf y}(\omega)$ is
\begin{align}\label{ytildeomega}
\tilde{\mathbf y}(\omega)={\mathbf F}\left({\mathbf y}\odot{\mathbf e}(-\omega)\right).
\end{align}
This means that
\begin{align}
\tilde{\mathbf y}(\omega)={\mathbf h}\odot {\mathbf d}+\tilde{\mathbf w}
\end{align}
is equivalent to the original observation model (\ref{simpfinal}), where $\tilde{\mathbf w}\sim {\mathcal {CN}}({\mathbf 0},\sigma^2{\mathbf I})$.

It is beneficial to introduce an additional hidden variable $\mathbf h$ defined in (\ref{hdef}), and an equivalent factor shown in Fig. \ref{FC_turbo_fig} (a) can be obtained. Compared to the original factor graph in Fig. \ref{FacOrig}, a delta factor node \cite{Mengunified} is introduced, which is the key to decompose the original problem into subproblems to be solved by designing the respective modules. According to Fig. \ref{FC_turbo_fig} (b), the JCCD-VALSE can be summarized as follows: Firstly, initialize the extrinsic message $m_{{\rm A}\rightarrow {\rm B}}({\mathbf h}_{\mathcal M})$ from module A to module B as
\begin{align}
m_{{\rm A}\rightarrow {\rm B}}({\mathbf h}_{{\mathcal M}})={\mathcal {CN}}({\mathbf h}_{{\mathcal M}};{\mathbf h}_{{\rm A},{\mathcal M}}^{\rm ext}, {\rm diag}({\mathbf v}_{{\rm A},{\mathbf h}_{{\mathcal M}}}^{\rm ext})),
\end{align}
and calculate the extrinsic message $m_{{\rm B}\rightarrow {\rm C}}({\mathbf d}_{\mathcal D})$ from module B to module C. Secondly, running the LDPC decoder algorithm in module C to obtain the posterior means and variances of ${\mathbf d}_{\mathcal D}$. Thirdly, update the residual CFO and noise variance estimate in module B. Fourthly, calculate the extrinsic message $m_{{\rm B}\rightarrow {\rm A}}({\mathbf h}_{\mathcal M})$ from module B to module A. Finally, calculate the extrinsic message $m_{{\rm A}\rightarrow {\rm B}}({\mathbf h}_{\mathcal M})$ from module A to module B, which closes the algorithm. The details of the above steps are described below.
\subsection{Calculate the Message $m_{{\rm B}\rightarrow {\rm C}}({\mathbf d}_{\mathcal D})$}\label{BtoCofd}
According to the belief propagation (BP), the message $m_{{\rm B}\rightarrow {\rm C}}({\mathbf d}_{\mathcal D})$ is
\begin{align}\label{messageBtoCofd}
m_{{\rm B}\rightarrow {\rm C}}({\mathbf d}_{\mathcal D})&\propto\int m_{{\rm A}\rightarrow {\rm B}}({\mathbf h})p({\mathbf y}|{\mathbf h},{\mathbf d}_{\mathcal D};\omega,\sigma^2){\rm d}{\mathbf h}\notag\\
&=p({\mathbf y}|{\mathbf d}_{\mathcal D};\omega,\sigma^2)=\prod\limits_{i\in {\mathcal D}}{\mathcal {CN}}(\tilde{y}_i(\omega);h_{{\rm A},i}^{\rm ext}d_i,|d_i|^2v_{{\rm A},{h}_i}^{\rm ext}).
\end{align}
For each $i\in\mathcal D$, $d_i$ is drawn from a finite alphabet set, and the PMF $m_{{\rm B}\rightarrow {\rm C}}({\mathbf d}_{\mathcal D})$ can be obtained by evaluating the above Gaussian PDF at the points of the symbol alphabet ${\mathbb A}_D$ followed by normalization.
\subsection{Calculate the Posterior Means and Variances of ${\mathbf d}_{\mathcal D}$ in LDPC Module}\label{LDPCSec}
The message $m_{{\rm B}\rightarrow {\rm C}}({\mathbf d}_{\mathcal D})$ is input to the LDPC module, where the SPA is applied. After several iterations where the change of the loglikelihood ratio of the bits is less than a threshold or the number of iterations exceeds the setting maximum, the beliefs of the data symbols are obtained. These beliefs are further used to calculate the posterior means $d_{{\rm C},i}^{\rm post}$ and variances $v_{{\rm C},d_i}^{\rm post}$ of the data.
\subsection{Residual CFO and Noise Variance Estimation}
The type II approach \cite{Mackay1992} and the EM algorithm are used to estimate the normalized residual CFO $\omega$ and noise variance $\sigma^2$, respectively. The type-II method is also termed as type-II maximum likelihood (ML) method \cite{Berger1985}, which aims to to maximize the marginal likelihood in Bayesian models. To estimate the residual CFO via type II approach, the marginal likelihood $p(\tilde{\mathbf y}(\omega);\omega)$ is obtained. Define
\begin{align}
{\mathbf z}={\mathbf h}\odot {\mathbf d}.
\end{align}
and model (\ref{model1}) can be formulated as
\begin{align}\label{model1}
\tilde{\mathbf y}(\omega)= {\mathbf z}+\tilde{\mathbf w},
\end{align}
where $\tilde{\mathbf w}\triangleq {\mathbf F}^{\rm H}({\mathbf w}\odot{\mathbf e}(-\omega))\sim {\mathcal {CN}}({\mathbf 0},\sigma^2{\mathbf I})$.
Straightforward calculation shows that the posterior means ${\mathbf z}_{\rm B}^{\rm post}$ and variances ${\mathbf v}_{{\rm B},{\mathbf z}}^{\rm post}$ of $\mathbf z$ are
\begin{subequations}\label{postz}
\begin{align}
{\mathbf z}_{\rm B}^{\rm post}&={\mathbf h}_{\rm A}^{\rm post}\odot{\mathbf d}_{\rm C}^{\rm post},\label{postmeanz}\\
{\mathbf v}_{{\rm B},{\mathbf z}}^{\rm post}&=|{\mathbf d}_{\rm C}^{\rm post}|^2\odot{\mathbf v}_{{\rm A},{\mathbf h}}^{\rm post}+|{\mathbf h}_{\rm A}^{\rm post}|^2\odot{\mathbf v}_{{\rm C},{\mathbf d}}^{\rm post}+{\mathbf v}_{{\rm A},{\mathbf h}}^{\rm post}\odot{\mathbf v}_{{\rm C},{\mathbf d}}^{\rm post},\label{postvarz}
\end{align}
\end{subequations}
where the posterior means and variances of $\mathbf h$ are obtained in module A, i.e., the VALSE module, see Subsection \ref{messageAtoB} (eq. (\ref{post_means_A}) and eq. (\ref{post_vars_A})).  Suppose that $\mathbf z$ follows Gaussian distribution with diagonal covariance matrix and independent of $\tilde{\mathbf w}$, i.e., ${\mathbf z}\sim {\mathcal {CN}}({\mathbf z};{\mathbf z}_{\rm B}^{\rm post},{\rm diag}({\mathbf v}_{{\rm B},{\mathbf z}}^{\rm post}))$. For the pseudo measurement $\tilde{\mathbf y}(\omega)$, it follows $\tilde{\mathbf y}(\omega)\sim {\mathcal {CN}}(\tilde{\mathbf y}(\omega);{\mathbf z}_{\rm B}^{\rm post},{\rm diag}({\mathbf v}_{{\rm B},{\mathbf z}}^{\rm post})+\sigma^2{\mathbf I})$. The type II ML estimation problem can be formulated as
\begin{align}\label{opt1omega}
\hat{\omega}=\underset{\omega}{\operatorname{argmin}}~(\tilde{\mathbf y}(\omega)-{\mathbf z}_{\rm B}^{\rm post})^{\rm H}({\rm diag}({\mathbf v}_{{\rm B},{\mathbf z}}^{\rm post})+\sigma^2{\mathbf I})^{-1}(\tilde{\mathbf y}(\omega)-{\mathbf z}_{\rm B}^{\rm post})\triangleq g(\omega).
\end{align}
One can apply the Newston step to refine the previous estimate $\hat{\omega}_g^{'}$ as
\begin{align}\label{NSToppler}
\hat{\omega}_g=\hat{\omega}_g^{'}-\frac{\partial g(\omega)}{\partial \omega}/\frac{\partial^2 g(\omega)}{\partial \omega^2} \Big|_{\omega=\hat{\omega}_g^{'}},
\end{align}
where
\begin{align}
\frac{\partial g(\omega)}{\partial \omega}=2\Re\left\{\left(\frac{\partial \tilde{\mathbf y}(\omega)}{\partial \omega}\right)^{\rm H}({\rm diag}({\mathbf v}_{{\rm B},{\mathbf z}}^{\rm post})+\sigma^2{\mathbf I})^{-1}(\tilde{\mathbf y}(\omega)-{\mathbf z}_{\rm B}^{\rm post})\right\}
\end{align}
and
\begin{align}
\frac{\partial^2 g(\omega)}{\partial \omega^2}&=2\Re\left\{\left(\frac{\partial^2 \tilde{\mathbf y}(\omega)}{\partial \omega^2}\right)^{\rm H}({\rm diag}({\mathbf v}_{{\rm B},{\mathbf z}}^{\rm post})+\sigma^2{\mathbf I})^{-1}(\tilde{\mathbf y}(\omega)-{\mathbf z}_{\rm B}^{\rm post})\right\}\notag\\
&+2\left(\frac{\partial \tilde{\mathbf y}(\omega)}{\partial \omega}\right)^{\rm H}({\rm diag}({\mathbf v}_{{\rm B},{\mathbf z}}^{\rm post})+\sigma^2{\mathbf I})^{-1}\frac{\partial \tilde{\mathbf y}(\omega)}{\partial \omega},
\end{align}
where
\begin{align}
\frac{\partial \tilde{\mathbf y}(\omega)}{\partial \omega}={\mathbf F}\left({\mathbf y}\odot{\boldsymbol \chi}\odot{\mathbf e}(-\omega)\right)
\end{align}
and
\begin{align}
\frac{\partial^2 \tilde{\mathbf y}(\omega)}{\partial \omega^2}={\mathbf F}\left({\mathbf y}\odot{\boldsymbol \chi}^2\odot{\mathbf e}(-\omega)\right),
\end{align}
where ${\boldsymbol \chi}=-{\rm j}[0,1,\cdots,N-1]^{\rm T}$. It is worth noting that the residual CFO estimation approach (\ref{opt1omega}) is very general and generalize the null subcarriers based approach in \cite{Zhou08JOE}. By letting ${\mathbf v}_{{\rm B},z_i}^{\rm post}\rightarrow \infty$, $i\in {\mathcal P}\cup {\mathcal D}$, i.e., the posterior variances of $\mathbf z$ corresponding to pilot and data subcarriers tend to infinity, (\ref{opt1omega}) reduces to the null subcarriers based approach \cite{Zhou08JOE}. Since we utilize more data to estimate the residual CFO, it makes sense that our approach estimates the residual CFO more accurately, leading to better channel estimation and data decoding performance, as illustrated in the numerical experiments.

Once the residual CFO $\omega$ is updated, the pseudo measurements $\tilde{\mathbf y}(\omega)$ are refined according to (\ref{ytildeomega}).

The EM approach is adopted to estimate the noise variance, which is
\begin{align}\label{EMobj}
\hat{\sigma}^2=\underset{{\sigma}^2}{\operatorname{argmax}}~{\rm E}_{{\mathbf z}}\left[\ln p(\tilde{\mathbf y}(\omega);{\mathbf z},\sigma^2)\right]=-\frac{{\rm E}_{{\mathbf z}}\|\tilde{\mathbf y}(\omega)-{\mathbf z}\|_2^2}{\sigma^2}-N\ln\sigma^2+{\rm const},
\end{align}
where the expectation is taken with respect to the posterior PDF of ${\mathbf z}$.

Substituting (\ref{postz}) in (\ref{EMobj}), one obtains
\begin{align}\label{EMobj1}
\hat{\sigma}^2=\underset{{\sigma}^2}{\operatorname{argmax}}~-\frac{\|\tilde{\mathbf y}(\omega)-{\mathbf z}_{\rm B}^{\rm post}\|_2^2+{\mathbf 1}^{\rm T}{\mathbf v}_{{\rm B},{\mathbf z}}^{\rm post}}{\sigma^2}-N\ln\sigma^2.
\end{align}
Setting the objective function of (\ref{EMobj1}) with respect to $\sigma^2$ to zero, the noise variance is estimated as
\begin{align}\label{EMobj1noisevar}
\hat{\sigma}^2=\frac{\|\tilde{\mathbf y}(\hat{\omega}_g)-{\mathbf z}_{\rm B}^{\rm post}\|^2+{\mathbf 1}^{\rm T}{\mathbf v}_{{\rm B},{\mathbf z}}^{\rm post}}{N}.
\end{align}

\subsection{Calculate the Extrinsic Message $m_{{\rm B}\rightarrow {\rm A}}({\mathbf h}_{\mathcal M})$ from Module B to Module A}\label{messageBtoA}
According to variational message passing \cite{VMP}, the message $m_{{\rm B}\rightarrow {\rm A}}({\mathbf h}_{\mathcal M})$ can be calculated as
\begin{align}\label{BAh}
\ln m_{{\rm B}\rightarrow {\rm A}}({\mathbf h}_{\mathcal M})&={\rm E}_{\mathbf d}\left[\ln p({\mathbf y}|{\mathbf h},{\mathbf d})\right]+{\rm const}\notag\\
&={\rm E}_{\mathbf d}\left[\ln p(\tilde{\mathbf y}(\omega)|{\mathbf h},{\mathbf d})\right]+{\rm const}\notag\\
&={\rm E}_{\mathbf d}\left[-\frac{|\tilde{\mathbf y}(\omega)-{\mathbf h}\odot{\mathbf d}|^2}{\sigma^2}\right]+{\rm const},
\end{align}
where ${\rm const}$ is to ensure that the PDF $m_{{\rm B}\rightarrow {\rm A}}({\mathbf h}_{\mathcal M})$ is normalized and ${\rm E}_{\mathbf d}[\cdot]$ denotes the expectation of $\mathbf d$ with respect to the posterior PDF. It can be seen that $m_{{\rm B}\rightarrow {\rm A}}({\mathbf h}_{\mathcal M})$ is Gaussian distributed.

From (\ref{BAh}), one has $m_{{\rm B}\rightarrow {\rm A}}({\mathbf h}_{\mathcal M})=\prod\limits_{i\in{\mathcal M}}m_{{\rm B}\rightarrow {\rm A}}({h}_i)$. Now we calculate the message $m_{{\rm B}\rightarrow {\rm A}}({h}_i)$ corresponding to the pilots and data subcarriers, respectively.

For $i\in {\mathcal P}$, straightforward calculation shows that
\begin{align}
m_{{\rm B}\rightarrow {\rm A}}({h}_i)={\mathcal {CN}}(h_i;h_{{\rm B},i}^{\rm ext},v_{{\rm B},h_i}^{\rm ext}),
\end{align}
where
\begin{align}\label{BApilot}
h_{{\rm B},i}^{\rm ext}=\tilde{y}_i(\omega)/d_i,\quad v_{{\rm B},h_i}^{\rm ext}=\sigma^2/|{d}_i|^2, i\in {\mathcal P}.
\end{align}

For the data symbols where $i\in {\mathcal D}$,
\begin{align}\label{BAhdata}
&\ln m_{{\rm B}\rightarrow {\rm A}}({h}_i)={\rm E}_{d_i}\left[\ln p(\tilde{\mathbf y}(\omega)|{\mathbf h},{\mathbf d})\right]+{\rm const}\notag\\
&={\rm E}_{\mathbf d}\left[-\frac{|\tilde{y}_i(\omega)-{h_i}d_i|^2}{\sigma^2}\right]+{\rm const}=-{\rm E}[|d_i|^2]\frac{|h|_i^2}{\sigma^2}+2\frac{\Re\{\tilde{y}_i^*(\omega)h_i{\rm E}[d_i]\}}{\sigma^2}+{\rm const}\notag\\
&=-\frac{(|d_{{\rm C},i}^{\rm post}|^2+v_{{\rm C},d_i}^{\rm post})|h|_i^2}{\sigma^2}+2\frac{\Re\{\tilde{y}_i^*(\omega)h_id_{{\rm C},i}^{\rm post}\}}{\sigma^2}+{\rm const}.
\end{align}
As a consequence, one has
\begin{subequations}\label{BAdata}
\begin{align}
h_{{\rm B},i}^{\rm ext}&=\frac{\tilde{y}_i(\omega)d_i^*}{|d_{{\rm C},i}^{\rm post}|^2+v_{{\rm C},d_i}^{\rm post}},\\
v_{{\rm B},h_i}^{\rm ext}&=\sigma^2/(|d_{{\rm C},i}^{\rm post}|^2+v_{{\rm C},d_i}^{\rm post}), i\in {\mathcal D}.
\end{align}
\end{subequations}
Note that if $v_{{\rm C},d_i}^{\rm post}=0$, (\ref{BAdata}) is consistent with (\ref{BApilot}).

\subsection{Calculate the message $m_{{\rm A}\rightarrow {\rm B}}({\mathbf h}_{\mathcal M})$}\label{messageAtoB}
Define ${\mathbf A}_{\mathcal M}$ as the matrix chosen from the rows of $\mathbf A$ indexed by ${\mathcal M}$. The pseudo linear measurement model under heterogenous noise is
\begin{align}
\widetilde{\mathbf h}_{\mathcal M}={\mathbf A}_{\mathcal M}({\boldsymbol \theta}){\boldsymbol \beta}+\widetilde{\boldsymbol \epsilon}_{\mathcal M},
\end{align}
where $\widetilde{\mathbf h}_{\mathcal M}={\mathbf h}_{{\rm B},{\mathcal M}}^{\rm ext}$, $\widetilde{\boldsymbol \epsilon}_{\mathcal M}\sim {\mathcal {CN}}({\mathbf 0},{\rm diag}(\tilde{\boldsymbol \sigma}_{\mathcal M}^2))$ and $\widetilde{\boldsymbol \sigma}_{\mathcal M}^2={\mathbf v}_{{\rm B},{\mathbf h}_{\mathcal M}}^{\rm ext}$.

The VALSE algorithm tries to construct a structured PDF $q({\boldsymbol \theta},{\boldsymbol \beta},{\mathbf s}|\widetilde{\mathbf h}_{\mathcal M})$ to approximate the true PDF $p({\boldsymbol \theta},{\boldsymbol \beta},{\mathbf s}|\widetilde{\mathbf h}_{\mathcal M})$ by minimizing their Kullback-Leibler (KL) divergence ${\rm{KL}}\left(q({\boldsymbol \theta},{\boldsymbol \beta},{\mathbf s}|\widetilde{\mathbf h}_{\mathcal M})||p({\boldsymbol \theta},{\boldsymbol \beta},{\mathbf s}|\widetilde{\mathbf h}_{\mathcal M})\right)$. For $q({\boldsymbol \theta},{\boldsymbol \beta},{\mathbf s}|\widetilde{\mathbf h}_{\mathcal M})$, it is supposed to be factored as \begin{align}
q({\boldsymbol \theta},{\boldsymbol \beta},{\mathbf s}|\widetilde{\mathbf h}_{\mathcal M}) = \prod_{i=1}^{L_{\rm max}}q(\theta_i|\widetilde{\mathbf h}_{\mathcal M})q({\boldsymbol \beta}|\widetilde{\mathbf h}_{\mathcal M},{\mathbf s})q({\mathbf s}|\widetilde{\mathbf h}_{\mathcal M}),\label{postpdf}
\end{align}
where $q({\mathbf s}|\widetilde{\mathbf h}_{\mathcal M})=\delta({\mathbf s}-\hat{\mathbf s})$ is restricted to a point estimate. Given that VALSE outputs the approximated PDF $q({\boldsymbol \theta},{\boldsymbol \beta},{\mathbf s}|\widetilde{\mathbf h}_{\mathcal M}) $, according to EP \cite{Minka}, the extrinsic message $m_{{\rm A}\rightarrow {\rm B}}({\mathbf h}_{\mathcal M})={\mathcal {CN}}({\mathbf h}_{\mathcal M};{\mathbf h}_{{\rm A},{\mathcal M}}^{\rm ext}, {\rm diag}({\mathbf v}_{{\rm A},{\mathbf h}_{\mathcal M}}^{\rm ext}))$ of ${\mathbf h}_{\mathcal M}$ from module A to module B is
\begin{align}\label{extA}
m_{{\rm A}\rightarrow {\rm B}}({\mathbf h}_{\mathcal M})&\propto\frac{{\rm Proj}[\int q({\boldsymbol \beta}_{\hat{S}}|\widetilde{\mathbf h}_{\mathcal M})\delta({\mathbf h}_{\mathcal M}-{\mathbf A}_{{\mathcal M},\hat{S}}({\boldsymbol \theta}){\boldsymbol \beta}_{\hat{S}})q({\boldsymbol \theta}|\widetilde{\mathbf h}_{\mathcal M}){\rm d}{\mathbf w}_{\hat{S}}{\rm d}{\boldsymbol \theta}]}{ m_{{\rm B}\rightarrow {\rm A}}({\mathbf h}_{\mathcal M})}\notag\\
&\triangleq \frac{{\rm Proj}[q_{\rm A}({\mathbf h}_{\mathcal M})]}{m_{{\rm B}\rightarrow {\rm A}}({\mathbf h}_{\mathcal M})},
\end{align}
where ${\rm Proj}[q({\mathbf x})]$ denotes the projection operation which approximates the target distribution $q({\mathbf x})$ as Gaussian distribution with moment matching, i.e., the means and variances of $q({\mathbf x})$ matches with that of the approximated Gaussian distribution. Let ${\mathbf h}_{{\rm A},{\mathcal M}}^{\rm post}$ and ${\mathbf v}_{{\rm B},{\mathbf h}_{\mathcal M}}^{\rm post}$ denote the posterior means and variances of ${\mathbf h}_{\mathcal M}$ with respect to $q_{\rm A}({\mathbf h}_{\mathcal M})$, then
\begin{align}\label{posthA}
{\rm Proj}[q_{\rm A}({\mathbf h}_{\mathcal M})]={\mathcal {CN}}({\mathbf h}_{\mathcal M};{\mathbf h}_{{\rm A},{\mathcal M}}^{\rm post}, {\rm diag}({\mathbf v}_{{\rm A},{\mathbf h}_{\mathcal M}}^{\rm post})).\end{align}
From (\ref{extA}), ${\mathbf h}_{{\rm A},{\mathcal M}}^{\rm ext}$ and ${\mathbf v}_{{\rm A},{\mathbf h}_{\mathcal M}}^{\rm ext}$ can be calculated as
\begin{subequations}
\begin{align}
&{\mathbf v}_{{\rm A},{\mathbf h}_{\mathcal M}}^{\rm ext}=\left(\frac{1}{{\mathbf v}_{{\rm A},{\mathbf h}_{\mathcal M}}^{\rm post}}-\frac{1}{{\mathbf v}_{{\rm A},{\mathbf h}_{\mathcal M}}^{\rm ext}}\right)^{-1},\label{extA_var}\\
&{\mathbf h}_{{\rm A},{\mathcal M}}^{\rm ext}={\mathbf v}_{{\rm A},{\mathbf h}_{\mathcal M}}^{\rm ext}\odot\left(\frac{{\mathbf h}_{{\rm A},{\mathcal M}}^{\rm post}}{{\mathbf v}_{{\rm A},{\mathbf h}_{\mathcal M}}^{\rm post}}-\frac{{\mathbf h}_{{\rm B},{\mathcal M}}^{\rm ext}}{{\mathbf v}_{{\rm B},{\mathbf h}_{\mathcal M}}^{\rm ext}}\right),\label{extA_mean}
\end{align}
\end{subequations}
where $\odot$ denotes the Hadarmard product. In the follow, the details of obtaining the approximated posterior PDFs through VALSE is briefly described \footnote{For further details, please refer to \cite{Badiu, ZZM18}.}.

The VALSE proceeds as follows: Firstly, the frequencies $\{\theta_i\}, i\in \hat{S}$ are inferred and the von Mises approximations $q(\theta_i|\widetilde{\mathbf h}_{\mathcal M}), i\in \hat{S}$ of the PDFs are obtained. Secondly, the weights $\boldsymbol \beta$ and support $\mathbf s$ are inferred and the posterior PDFs $q({\boldsymbol \beta}|\widetilde{\mathbf h}_{\mathcal M})$ of the weight and the PMF $q({\mathbf s}|\widetilde{\mathbf h}_{\mathcal M})$ of the support are obtained. Finally, the model parameters $\rho$ and $\nu$ are estimated. In the following, we detail the procedures.

\subsubsection{Inferring the frequencies}
Let $\mathcal S$ be the set of indices of the non-zero components of $\mathbf s$, i.e.,
\begin{align}\notag
\mathcal S = \{i|1\leq i\leq L_{\rm max},s_i = 1\}.
\end{align}
Analogously, define $\widehat{\mathcal S}$ based on $\widehat{\mathbf s}$. For $i\notin{\mathcal S}$, we have $q(\theta_i|\widetilde{\mathbf h}_{\mathcal M})= p(\theta_i)$. For $i\in{\mathcal S}$, $q(\theta_i|\widetilde{\mathbf h}_{\mathcal M})$ is
\begin{align}\label{pdf-q}
q(\theta_i|\widetilde{\mathbf h}_{\mathcal M})\propto p(\theta_i){\rm exp}(\Re\{{\boldsymbol\eta}_i^{\rm H}{\mathbf a}(\theta_i)\})\triangleq \exp\{f(\theta_i)\},
\end{align}
where the complex vector $\boldsymbol\eta_i$ is given by
\begin{align}\label{yita-i}
{\boldsymbol\eta}_i = 2\left[\left(\widetilde{\mathbf h}_{\mathcal M}-\sum_{l\in\widehat{\mathcal S} \backslash \{i\}}{\widehat{\mathbf a}}_{{\mathcal M},l}{\widehat{\beta}}_l\right){\widehat{\beta}}^*_i-\sum_{l\in\widehat{\mathcal S} \backslash \{i\}}{\widehat{\mathbf a}}_{{\mathcal M},l}{\widehat{C}}_{l,i}\right]/\tilde{\boldsymbol \sigma}_{\mathcal M}^2,
\end{align}
where ${\widehat{\mathbf a}}_{{\mathcal M},l}$ denotes the $l$th column of ${\widehat{\mathbf A}}_{{\mathcal M}}$, $``\backslash i"$ denotes the indices $\widehat{\mathcal S}$ excluding $i$, $\hat{\boldsymbol \beta}_{\widehat{\mathcal S}}$ denotes the subvector of $\hat{\boldsymbol \beta}$ by choosing the $\widehat{\mathcal S}$ rows of ${\boldsymbol \beta}$.  Since it is hard to obtain the analytical results of the expected value of ${\mathbf a}_{\mathcal M}(\theta_i)$ for the PDF (\ref{pdf-q}), $q(\theta_i|\tilde{\mathbf h})$ is approximated as a von Mises distribution via the following two steps \footnote{For further details, please refer to \cite[Algorithm 2: Heurestic 2]{Badiu}.}: Firstly, search for the most dominant component of (\ref{pdf-q}) via Heuristic 2 \cite[Algorithm 2]{Badiu}. Secondly, use second-order Taylor approximation to approximate the dominant component as a single von Mises PDF. It is worth noting that, the number of subcarriers $N$ is usually very large, for example $N=1024$, thus the peak of the single von Mises PDF should be well matched to that of (\ref{pdf-q}) in very high accuracy. Let $\bar{\theta}_i$ denote the mean of the dominant component. In contrast with \cite{Badiu} where a single Newston refinement is adopted, we apply a two step procedure to improve the robustness. First, apply the Newston step to refine the estimate
\begin{align}
\hat{\mu}_i=\bar{\theta}_i-\frac{f^{'}(\bar{\theta}_i)}{f^{''}(\bar{\theta}_i)},
\end{align}
where $f(\theta_i)$ is defined in (\ref{pdf-q}). Since the Newston step may be too large to across the peak, we set
\begin{align}
\bar{\mu}_i=\frac{\hat{\mu}_i+\bar{\theta}_i}{2}.
\end{align}
Then we continue to apply the Newston step to obtain the final estimate as
\begin{align}
\hat{\theta}_i=\bar{\mu}_i-\frac{f^{'}(\bar{\mu}_i)}{f^{''}(\bar{\mu}_i)}.
\end{align}
Numerical experiments demonstrate that such a two step approach improves the robustness of the VALSE significantly, compared to the original implementation \cite{Badiu}. Besides, the concentration parameter $\hat{\kappa}_i$ is approximated as
\begin{align}
\hat{\kappa}_i=A^{-1}\left({\rm e}^{0.5/f^{''}(\hat{\theta}_i)}\right),
\end{align}
where $A^{-1}(\cdot)$ is the inverse function of $A(\cdot)=I_1(\cdot)/I_0(\cdot)$. The expected value of ${\mathbf a}_{\mathcal M}(\theta_i)$ is
\begin{align}
\hat{\mathbf a}_{\mathcal M}(\theta_i)=\frac{I_{\mathcal M}(\hat{\kappa}_i)}{I_0(\hat{\kappa}_i)}\odot{\mathbf a}_{\mathcal M}(\hat{\theta}_i).
\end{align}

It is found that calculating the most dominate component via Heurestic 2 \cite[Algorithm 2]{Badiu} in each iteration is very time consuming. Consequently, we adopt the warm start strategy where we only implement Heuristic 2 in the initialization stage. Then we use the previous estimates to perform refinement, which is very efficiently.
\subsubsection{Inferring the weights and support}
Next by fixing the posterior PDFs $q(\theta_i|\widetilde{\mathbf h}_{\mathcal M}),i=1,...,L_{\rm max}$, the posterior PDF $q({\boldsymbol \beta},{\mathbf s}|\tilde{\mathbf h}_{\mathcal M})$ is obtained. Define the matrices $\mathbf J$ and $\mathbf u$ as
\begin{subequations}\label{J-H}
\begin{align}
&{J}_{ij}= 	
\begin{cases}
{\mathbf 1}_{\mathcal M}^{\rm T}\left(\frac{1}{\tilde{\boldsymbol \sigma}_{\mathcal M}^2}\right),&i=j\\
{\widehat{\mathbf a}}^{\rm H}_{{\mathcal M},i}{\rm diag}\left(\frac{1}{\tilde{\boldsymbol \sigma}_{\mathcal M}^2}\right){\widehat{\mathbf a}}_{{\mathcal M},j},&i\neq{j}
\end{cases},\quad i,j\in\{1,2,\cdots,L_{\rm max}\},\label{J-H1}\\
&{\mathbf u} = \widehat{\mathbf A}_{\mathcal M}^{\rm H}\left(\tilde{\mathbf h}_{\mathcal M}/\tilde{\boldsymbol \sigma}_{\mathcal M}^2\right).\label{J-H2}
\end{align}
\end{subequations}

According to (\ref{postpdf}), the posterior approximation $q({\boldsymbol \beta},{\mathbf s}|\tilde{\mathbf h}_{\mathcal M})$  is factored as the product of $q({\boldsymbol \beta}|\tilde{\mathbf h}_{\mathcal M},{\mathbf s})$ and $\delta({\mathbf s}-{\widehat{\mathbf s}})$. For a given $\widehat{\mathbf s}$, $q({\boldsymbol \beta}_{\widehat{\mathcal S}}|\tilde{\mathbf h}_{\mathcal M})$ is a complex Gaussian distribution, and $q({\boldsymbol \beta}|\tilde{\mathbf h}_{\mathcal M};\widehat{\mathbf s})$ is
\begin{align}
q({\boldsymbol \beta}|\tilde{\mathbf h};\widehat{\mathbf s}) = {\mathcal {CN}}({\boldsymbol \beta}_{\widehat{\mathcal S}};\widehat{\boldsymbol \beta}_{\widehat{\mathcal S}},\widehat{\mathbf C}_{\widehat{\mathcal S}})\prod_{i\not\in\widehat{\mathcal S}}\delta(\beta_{i}),\label{qwpdfiid}
\end{align}
where
\begin{subequations}\label{W-C-1}
\begin{align}
&\widehat{\mathbf C}_{{\mathcal S}} = \left({{\mathbf J}_{{\mathcal S}}}+\frac{{\mathbf I}_{|{\mathcal S}|}}{\nu}\right)^{-1},\label{Covcal}\\
&\widehat{\boldsymbol \beta}_{{\mathcal S}} = \widehat{\mathbf C}_{{\mathcal S}}{\mathbf u}_{{\mathcal S}}.\label{What}
\end{align}
\end{subequations}
Although (\ref{Covcal}) involves a matrix inversion, it can be avoided by utilizing the rank one update formula.

Then we need to find $\hat{\mathbf s}$ which maximizes $\ln Z(\mathbf s)$, i.e.,
\begin{align}\label{maxlns}
\hat{\mathbf s}=\underset{\mathbf s}{\operatorname{argmax}}\ln Z(\mathbf s),
\end{align}
where $\ln Z(\mathbf s)$ is
\begin{align}\label{L-ws}
\ln Z(\mathbf s)\triangleq&-\ln\det\left({\mathbf J}_{\mathcal S}+\frac{1}{\nu}{\mathbf I}_{|\mathcal S|}\right)+{\mathbf u}_{\mathcal S}^{\rm H}\left(\mathbf J_{\mathcal S}+\frac{1}{{\nu}}\mathbf I_{|\mathcal S|}\right)^{-1}{\mathbf u}_{\mathcal S}\notag\\
&+||\mathbf s||_0\ln\frac{\rho}{1-\rho}+||\mathbf s||_0\ln\frac{1}{{\nu}}+{\rm const}.
\end{align}
Problem (\ref{maxlns}) is solved via a greedy algorithm, and local optimum is guaranteed.
\subsubsection{Inferring the model parameters}
After updating the PDFs of the frequencies and weights, the model parameters $\{\rho,~\nu\}$ are updated as
\begin{align}\label{mu-rou-tau-hat}
\widehat{\rho} = &\frac{||\widehat{\mathbf s}||_0}{{L_{\rm max}}},\notag\\
\widehat{{\nu}} = &\frac{\widehat{\boldsymbol \beta}^{\rm H}_{\widehat{\mathcal S}}\widehat{\boldsymbol \beta}_{\widehat{\mathcal S}}+{\rm tr}(\widehat{{\mathbf C}}_{\widehat{\mathcal S}})}{||\widehat{\mathbf s}||_0}.
\end{align}

Now the the posterior mean and variances of ${\mathbf h}$ can be obtained as
\begin{subequations}\label{post_meanvar_h}
\begin{align}
&{\mathbf h}_{{\rm A},{\mathcal M}}^{\rm post}=\hat{\mathbf A}_{{\mathcal M},\hat{\mathcal S}}\hat{\boldsymbol \beta}_{\hat{\mathcal S}},\label{post_means_A}\\
&{\mathbf v}_{{\rm A},{\mathbf h}_{\mathcal M}}^{\rm post}= {\rm diag}(\hat{\mathbf A}_{{\mathcal M},\hat{\mathcal S}}\hat{\mathbf C}_{\hat{\mathcal S}}\hat{\mathbf A}_{{\mathcal M},\hat{\mathcal S}}^{\rm H})+\left(\hat{\boldsymbol \beta}_{\hat{\mathcal S}}^{\rm H}\hat{\boldsymbol \beta}_{\hat{\mathcal S}}{\mathbf 1}_{|{\mathcal M}|}-|\hat{\mathbf A}_{{\mathcal M},\hat{\mathcal S}}|^2|\hat{\boldsymbol \beta}_{\hat{\mathcal S}}|^2\right)\notag\\
&+\left[{\rm tr}(\hat{\mathbf C}_{\hat{\mathcal S}}){\mathbf 1}_{|{\mathcal M}|}-|\hat{\mathbf A}_{{\mathcal M},\hat{\mathcal S}}|^2{\rm diag}(\hat{\mathbf C}_{\hat{\mathcal S}})\right],\label{post_vars_A}
\end{align}
\end{subequations}
respectively. Then, the extrinsic message $m_{{\rm A}\rightarrow {\rm B}}({\mathbf h}_{\mathcal M})={\mathcal {CN}}({\mathbf h}_{\mathcal M};{\mathbf h}_{{\rm A},{\mathcal M}}^{\rm ext},{\rm diag}({\mathbf v}_{{\rm A},{\mathbf h}_{\mathcal M}}^{\rm ext}))$ from module A to module B can be obtained from (\ref{extA_mean}) and (\ref{extA_var}). Remember that we have initialized the extrinsic message $m_{{\rm A}\rightarrow {\rm B}}({\mathbf h}_{\mathcal M})$ from module A to module B at the beginning of Section \ref{Algorithm}, the algorithm is closed.
\subsection{Initialization}\label{Init}
Performing the joint estimation of the channel, residual CFO and data detection is a very challenging task, and initialization is a very important step to have good performance. Since the residual CFO $\omega$ is close to zero, we initialize it as zero. To obtain a good estimate of the channel, we run the pilot based VALSE algorithm to obtain the posterior means and variances of the channel, which are provided as the initialization of the extrinsic message $m_{{\rm A}\rightarrow {\rm B}}({\mathbf h}_{\mathcal M})$ from module A to module B. Besides, we also initialize the nuisance parameters $\sigma^2$ and ${\nu}$ via the pilot based VALSE algorithm. The whole algorithm is summarized as Algorithm \ref{JCDVALSE}.
\begin{algorithm}[h]
\caption{ JCCD-VALSE algorithm}\label{JCDVALSE}
\begin{algorithmic}[1]
\STATE Initialize the extrinsic message $m_{{\rm A}\rightarrow {\rm B}}({\mathbf h}_{\mathcal M})$ from module A to module B, the nuisance parameters $\sigma^2$ and $\nu$ via the pilot based VALSE algorithm, as described in Subsection \ref{Init}; Set the number of iterations $T_{\rm max}$, the maximum number of paths $L_{\rm max}$;
\STATE Calculate the message $m_{{\rm B}\rightarrow {\rm C}}({\mathbf d}_{\mathcal D})$ (\ref{messageBtoCofd}) and input it to the LDPC decoder module to obtain the posterior means $d_{{\rm C},i}^{\rm post}$ and variances $v_{{\rm C},d_i}^{\rm post}$ of the data, as described in Subsection \ref{LDPCSec}.
\STATE Calculate the extrinsic message $m_{{\rm B}\rightarrow {\rm A}}({\mathbf h}_{\mathcal M})$ through (\ref{BApilot}) and (\ref{BAhdata}) from module B to module A.
\STATE Initialize $q(\theta_i|\widetilde{\mathbf h}_{\mathcal M})$, $i=1,2,\cdots,L_{\rm max}$; compute $\hat{\mathbf a}_{\mathcal M}(\theta_i)$, $\mathbf J$ (\ref{J-H1}) and $\mathbf u$ (\ref{J-H2}).
\FOR {$t=1,\cdots,T_{\rm max}$ }
\STATE Update~$\widehat{\mathbf s}$ (\ref{maxlns}), $\widehat{\boldsymbol \beta}_{\widehat{\mathcal S}}~{\rm and}~\widehat{\mathbf C}_{\widehat{\mathcal S}}$ (\ref{W-C-1}).
\STATE Update~$\widehat{\rho}$, $\widehat{\nu}$ (\ref{mu-rou-tau-hat}) and Update~$\boldsymbol\eta_i$~and~$\widehat{\mathbf a}_i$ for all $i\in \widehat{\mathcal S}$.
\STATE Compute the posterior means and variances of $\mathbf h$ as ${\mathbf h}_{\rm A}^{\rm post}$ (\ref{post_means_A}), ${\mathbf v}_{{\rm A},{\rm h}}^{\rm post}$ (\ref{post_vars_A}).
\STATE Calculate the extrinsic message $m_{{\rm A}\rightarrow {\rm B}}({\mathbf h}_{\mathcal M})$ from module B to module A.
\STATE Calculate the message $m_{{\rm B}\rightarrow {\rm C}}({\mathbf d}_{\mathcal D})$ (\ref{messageBtoCofd}) and input it to the LDPC decoder module to obtain the posterior means $d_{{\rm C},i}^{\rm post}$ and variances $v_{{\rm C},d_i}^{\rm post}$ of the data.
\STATE Compute the posterior means and variances of $\mathbf z$ as ${\mathbf z}_{\rm B}^{\rm post}$ (\ref{postmeanz}) and ${\mathbf v}_{{\rm B},{\mathbf z}}^{\rm post}$ (\ref{postvarz}), respectively. Implement the Newston step to update the residual CFO and the EM step to update the noise variance via (\ref{NSToppler}) and (\ref{EMobj1noisevar}), respectively.
\STATE Calculate the extrinsic message $m_{{\rm B}\rightarrow {\rm A}}({\mathbf h}_{\mathcal M})$ through (\ref{BApilot}) and (\ref{BAhdata}) from module B to module A.
\STATE Update $\mathbf J$ (\ref{J-H1}) and $\mathbf u$ (\ref{J-H2}).
\ENDFOR
\STATE Return belief function of data bits $u_k$, $k\in {\mathcal K}$, channel estimate $\hat{\mathbf h}$.
\end{algorithmic}
\end{algorithm}

\section{Numerical Simulation Results}\label{NSres}
In this section, numerical simulations are conducted to demonstrate the advantages of the proposed joint processing scheme. A typical underwater acoustic channel setting is adopted in the simulation: there are $15$ paths, the arrival time between two adjacent paths follows exponential distribution with mean value $1$ ms, and the amplitude of each path follows Rayleigh distribution with mean power decreasing by $20$ dB in $30$ ms delay spread~\cite{bzpw10}. The simulation parameters of the OFDM system are listed in Table~\ref{tab.ofdm_para}. In addition, a common Doppler scaling factor exists in all paths, which follows uniform distribution in the range $\left[ -2\times 10^{-3}, 2\times 10^{-3}   \right]$. In our simulation, null subcarriers are utilized to estimate the Doppler scaling factor and the residual CFO~\cite{wwzy12}, based on the results of which the resampling and CFO compensation are implemented. Performance metrics of decoding BER and NMSE of channel estimation defined as
\begin{align}\label{equ.chan_nmse}
{\rm NMSE}(\hat{\mathbf h}) = 10\log_{10}\left(\frac{\left\| \hat{\mathbf h } - \mathbf{h}_0 \right\|^2}{\left\|\mathbf{h}_0 \right\|^2}\right)\notag
\end{align}
are compared in the simulation, where $\mathbf{h}_0$, $\hat{\mathbf{h}}$ are the ground truth and estimated value of the frequency-domain channel response vector, respectively. All the results are averaged over $500$ MC trials.

{\color{red}
\begin{table}[h]
\renewcommand\arraystretch{1.2}
  \centering
  \footnotesize
      \caption{CP-OFDM parameter settings}\label{tab.ofdm_para}
    \begin{tabular}{|ll|l|}
\hline
      Bandwidth           & $B$    & $4882.8~\text{Hz}$ \\\hline
      Carrier frequency   & $f_c$  & $13000~\text{Hz}$ \\\hline
      Sampling frequency  & $f_s$  & $39062.5~\text{Hz}$ \\\hline
      No. subcarriers      & $N$    & $1024$ \\\hline
      No. data subcarriers & $N_{\mathcal D}$  & $672$ \\\hline
      No. pilot subcarriers& $N_{\mathcal P}$   & $256$ \\\hline
      No. null subcarriers& $N_{\mathcal N}$   & $96$ \\\hline
      Symbol duration     & $T$    & $209.72~\text{ms}$ \\\hline
      Cyclic-prefix length& $T_\text{cp}$ & $40.35~\text{ms}$ \\
      \hline
    \end{tabular}
    \normalsize
\end{table}
}

\subsection{Benchmark Algorithms}
Several approaches classified as pilot-only based and joint channel and data decoding based are implemented to make performance comparison.
The pilot-only based are listed as follows:
\begin{itemize}
  \item The pilot based OMP approach \cite{bzpw10}. The dictionary is constructed by $\times 8$ oversampling for time delay search, and the Bayesian information stopping criterion \cite{OMPBIC} is adopted, and the algorithm is termed as OMP.
  \item The pilot based AMP approach \cite{EMAMPconf}. It is numerically found that AMP diverges with the oversampling factor greater than $2$. We construct the dictionary by $\times 2$ oversampling. Besides, the Bernoulli Gaussian prior is adopted for the coefficients and the EM algorithm \cite{EMAMPconf, EMAMP} is incorporated to jointly estimate the nuisance parameters of the prior distribution and the noise variance. The algorithm is termed as AMP.
  \item The pilot based VALSE approach \cite{Badiu}. The observations corresponding to pilots are directly input to the VALSE algorithm to obtain the path delays and amplitudes, which are then used to reconstruct the whole channel. The algorithm is termed as VALSE.
\end{itemize}
Note that after the whole channel is reconstructed, the estimates (including means and variances) of the data are input to the single input single output (SISO) equalizer, i.e, iterative decoding is conducted to obtain the BER.

The GAMP based \emph{j}oint \emph{c}hannel estimation and \emph{d}ata decoding approach (JCD) named JCD-GAMP is also conducted \cite{SchniterJAMP}. Note that the original approach uses the Markov chain prior and learn the nuisance parameters of the prior via fitting the simulated channel data. Besides, it uses the true noise variance. Here we use the Bernoulli Gaussian prior and use the EM to jointly estimate the nuisance parameters of the prior distribution and the noise variance instead.

To show the benefits of estimating the residual CFO, we also implement the VALSE based JCD approach without estimating the CFO, and we name the approach as JCD-VALSE.

To further show the effectiveness of the proposed algorithm, we also evaluate the coded BER under the perfect channel state information (PCSI) and the NMSE of the JCCD-VALSE with data aware, which are termed as PCSI and JCCD-VALSE (data aware), respectively. Note that the coded BER of PCSI and the JCCD-VALSE (data aware) will be the lower bounds of the coded BER and the NMSE, respectively.

\subsection{QPSK Results}\label{BER_res}
\begin{figure*}
  \centering
  \subfigure[]{
    \label{QPSK_BER_fig} 
    \includegraphics[width=65mm]{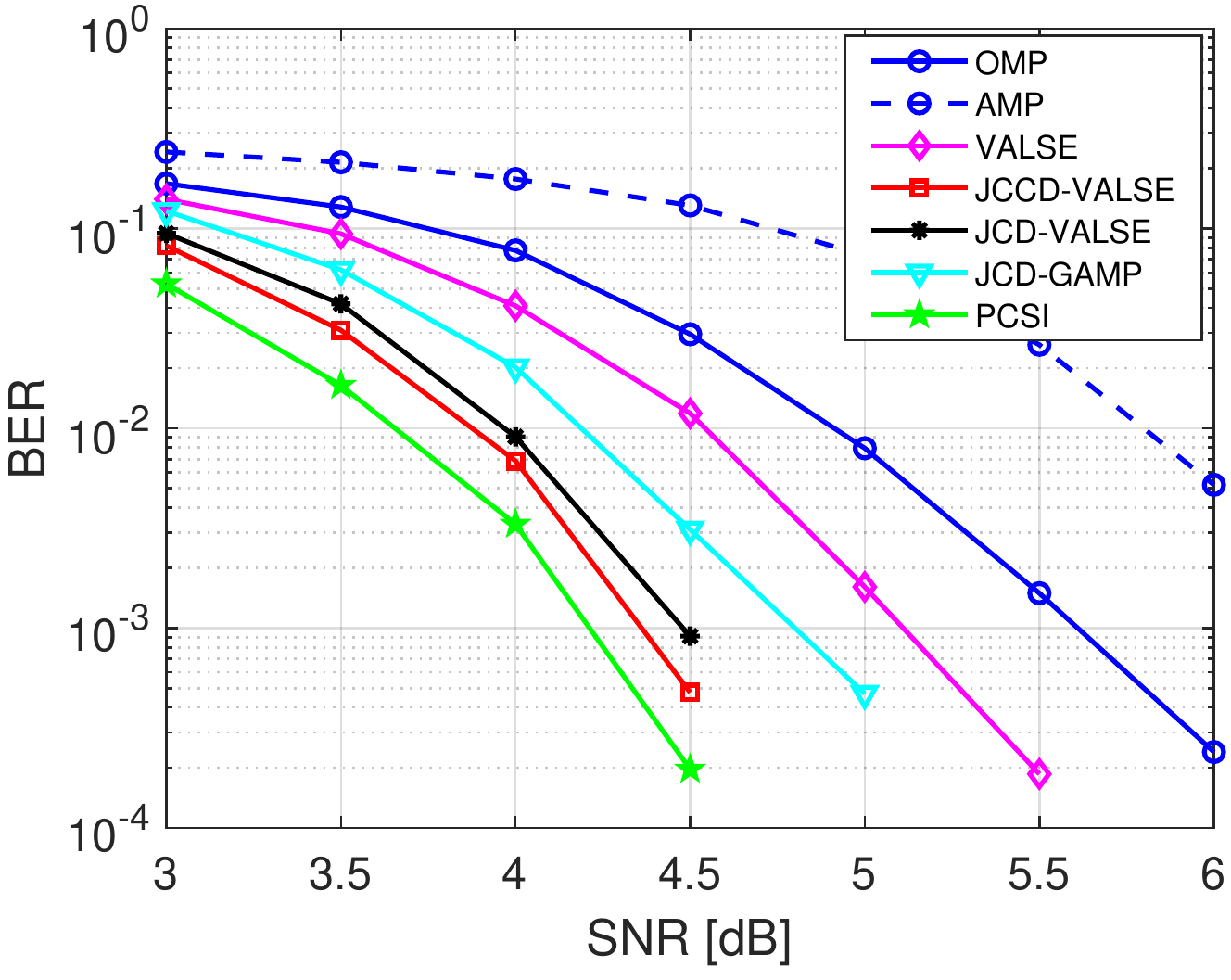}}
        \hspace{0.2in}
    \subfigure[]{
    \label{QPSK_NMSE_fig}
    \includegraphics[width=65mm]{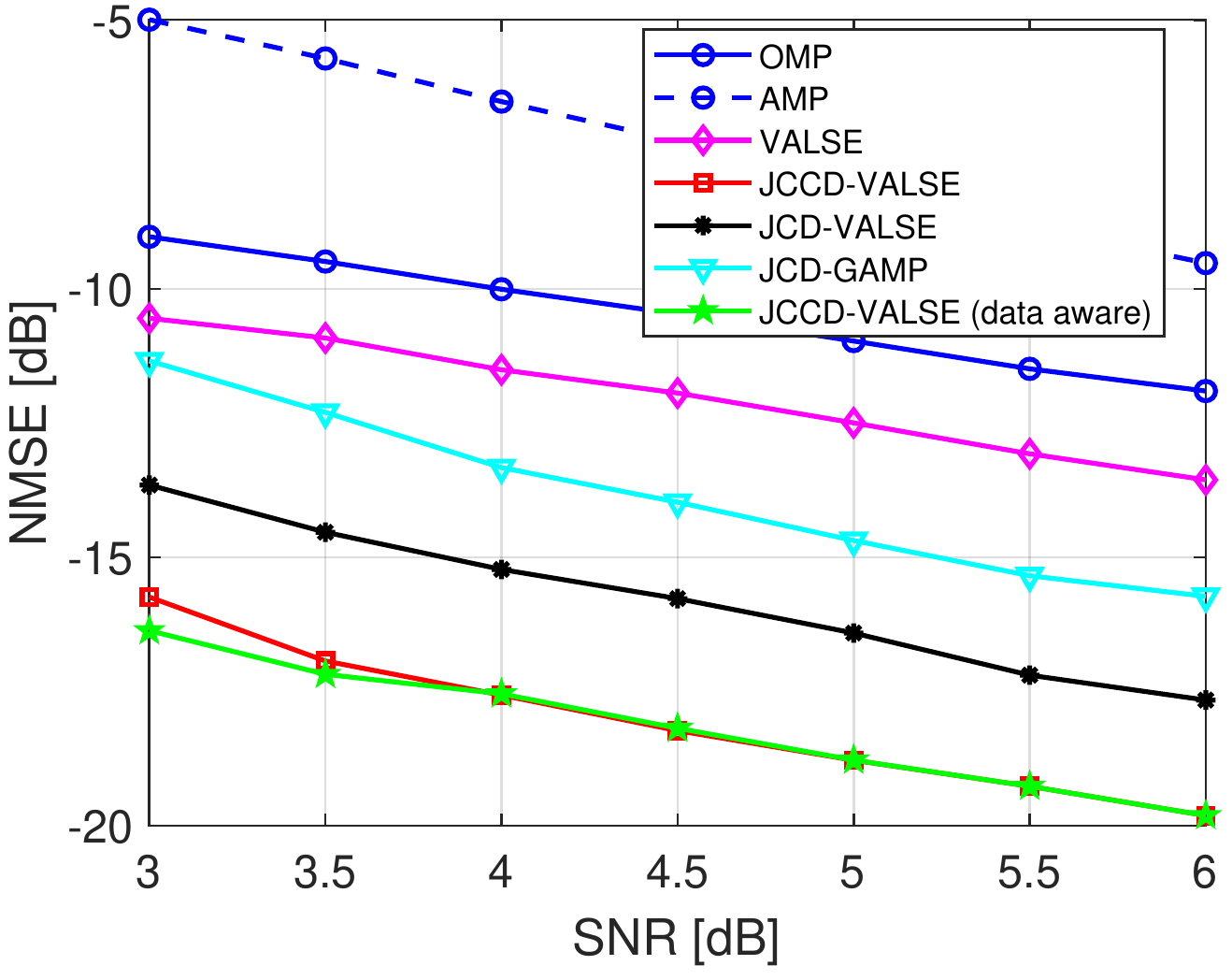}}
  \caption{Performance metrics evaluated by the various algorithms for QPSK. (a) coded BER, (b) NMSE.}
  \label{QPSK_all_fig} 
\end{figure*}

Fig.~\ref{QPSK_all_fig} presents the performance metrics evaluated by the various algorithms for QPSK modulation. For the coded BER shown in Fig. \ref{QPSK_BER_fig}, PCSI works best, followed by JCCD-VALSE, JCD-VALSE, JCD-GAMP, VALSE, OMP and AMP. For ${\rm BER}=10^{-3}$, the SNR needed for the PCSI, JCCD-VALSE, JCD-VALSE, JCD-GAMP, VALSE, and OMP are $4.21$ dB, $4.36$ dB, $4.47$ dB, $4.81$ dB, $5.11$ dB, $6.61$ dB. The SNR gap between the JCCD-VALSE and the PCSI method (oracle) is $4.36-4.21=0.15$ dB, which is very close. Compared to JCD-GAMP, VALSE and AMP approaches, our JCCD method has SNR gains $4.81-4.36=0.45$ dB, $5.11-4.36=0.75$ dB, $6.61-4.36=1.75$ dB, respectively. As for the NMSE of the channel estimation, Fig. \ref{QPSK_NMSE_fig} shows that JCCD-VALSE (data aware) works best, followed by JCCD-VALSE, JCD-VALSE (without estimating the residual CFO), JCD-GAMP, VALSE, OMP and AMP. Besides, JCCD-VALSE asymptotically approaches to JCCD-VALSE (data aware). To sum up, the proposed approach achieves excellent data decoding and channel estimation performance for QPSK modulated symbols.

\subsection{$16$ QAM Results}
\begin{figure*}
  \centering
  \subfigure[]{
    \label{16QAM_BER_fig} 
    \includegraphics[width=65mm]{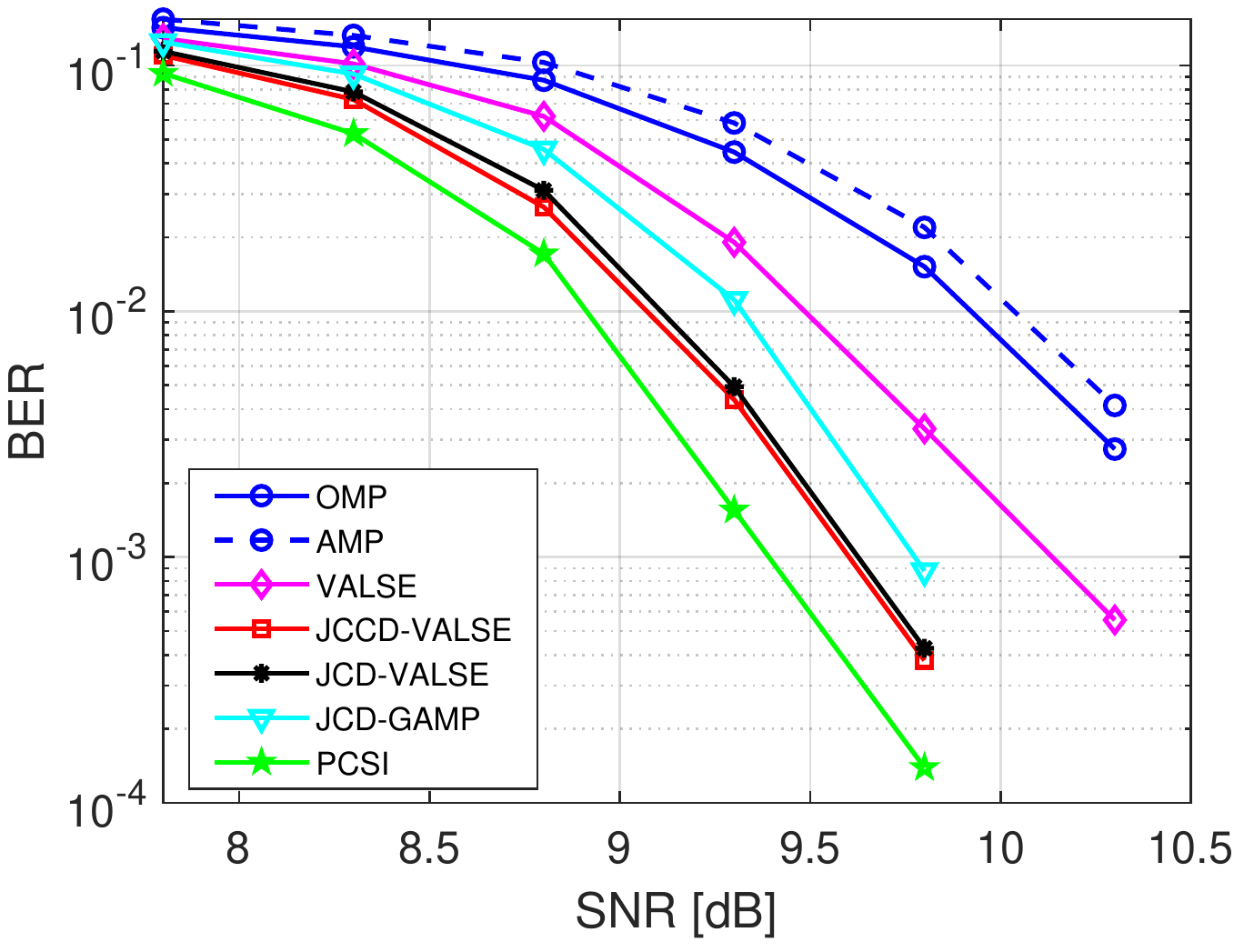}}
        \hspace{0.2in}
    \subfigure[]{
    \label{16QAM_NMSE_fig}
    \includegraphics[width=65mm]{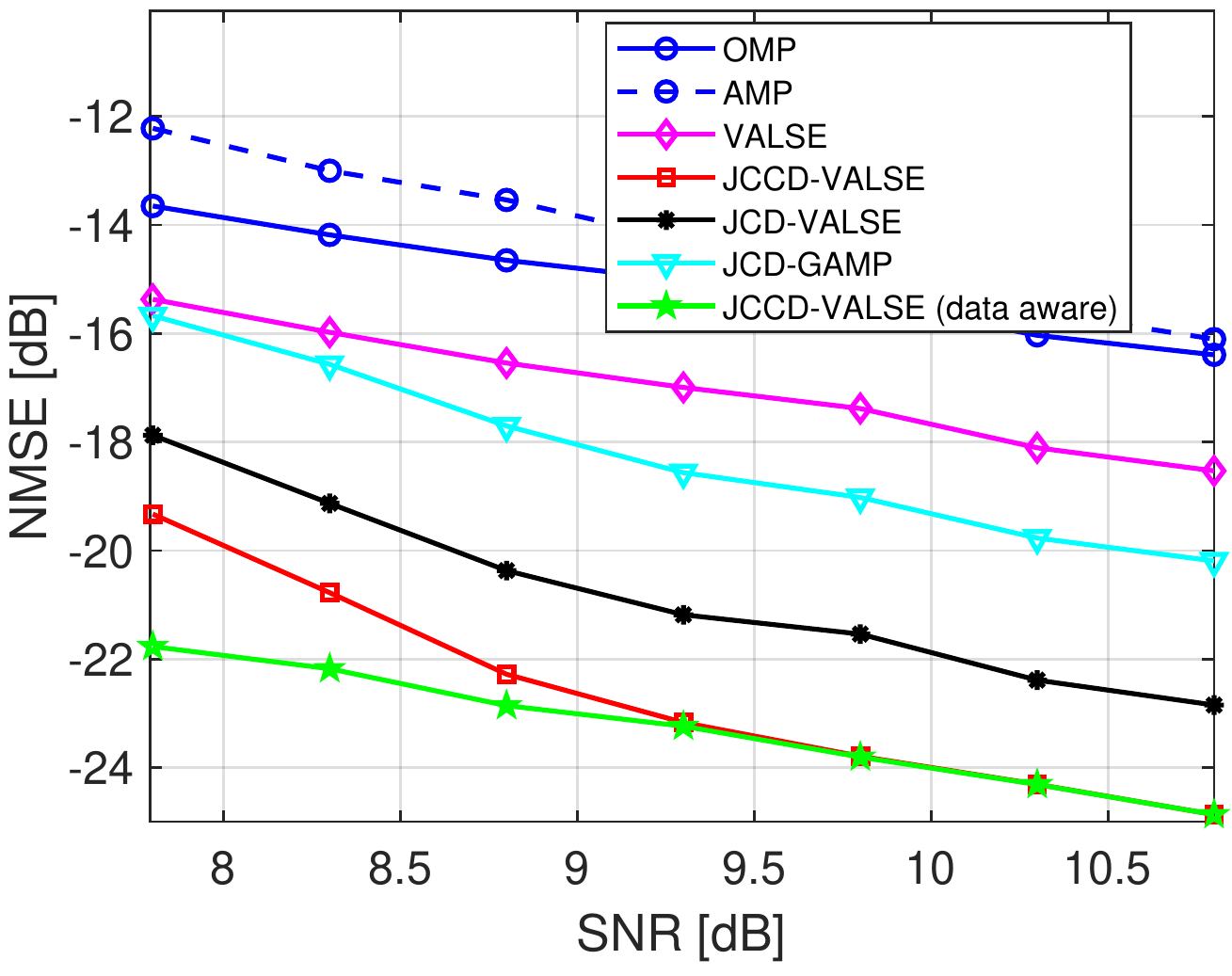}}
  \caption{Performance metrics evaluated by the various algorithms for $16$ QAM. (a) coded BER, (b) NMSE.}
  \label{QAM16_all_fig} 
\end{figure*}

The performance metrics evaluated by the various algorithms for $16$ QAM modulation are shown in Fig.~\ref{QAM16_all_fig}. Note that the results are similar to the QPSK modulation setting except the following two differences. The first difference is that because the constellation becomes larger, the SNR needed to achieve the same coded BER and the NMSE increases (from about $4.36$ dB to $9.7$ dB of JCCD-VALSE). The second difference is that the gain of the CFO estimation for the JCCD-VALSE is marginal, compared to JCD-VALSE (without estimating the residual CFO). The reason is probably that the Doppler scaling factor estimation accuracy is better in the higher SNR region, which leads to smaller residual CFO. Still, we conclude that the proposed approach has performance gains compared to the other approaches for $16$ QAM modulated symbols.

\section{Sea Trial Data Decoding Results}\label{Realdata}

\begin{figure*}
  \centering
  \subfigure[Estimated channel impulse response]{
    \label{16QAM_BER_fig} 
    \includegraphics[width=65mm]{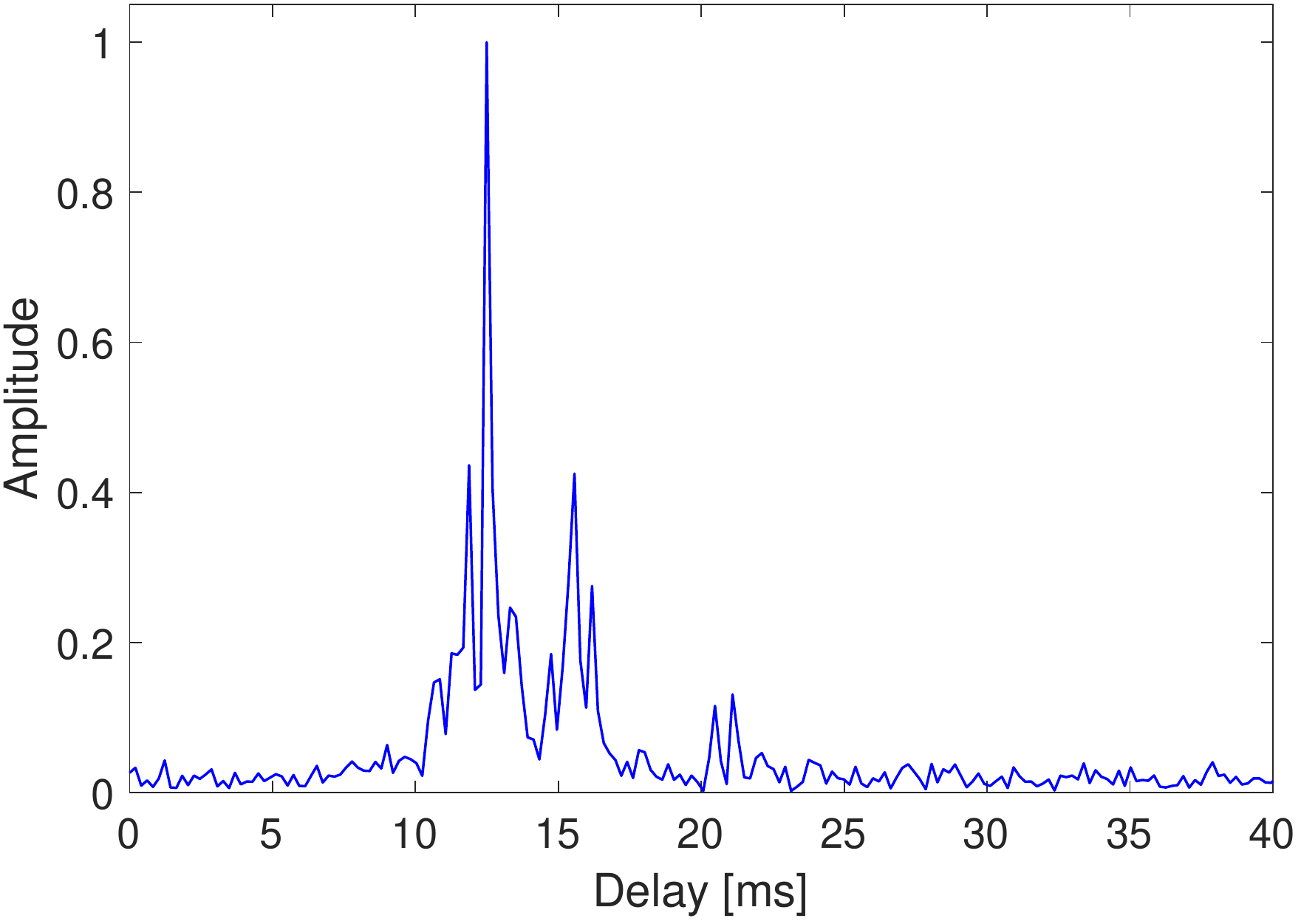}}
    \hspace{0.2in}
    \subfigure[Estimated Doppler speed]{
    \label{16QAM_NMSE_fig}
    \includegraphics[width=70mm]{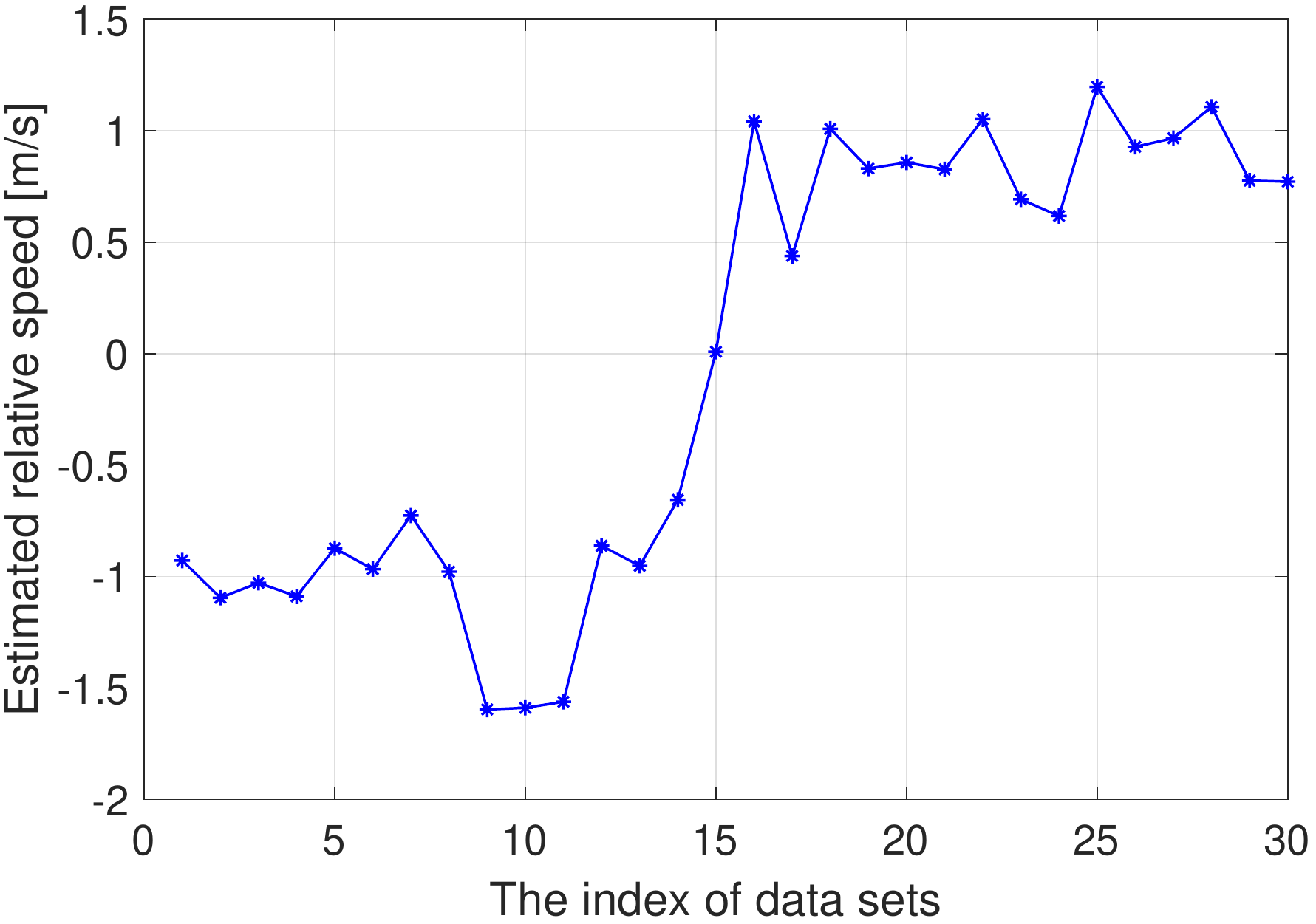}}
  \caption{Channel condition in MACE10~\cite{wjzm20}.}
  \label{mace10_chan_fig} 
\end{figure*}

To further verify the performance of the proposed JCCD-VALSE scheme in real underwater acoustic communication channels, sea test data from MACE10 is adopted for decoding. MACE10 experiment was carried out off the coast of Marthas Vineyard, Massachusetts, at Jun. 2010. The depth of water was about 80 meters. The source was towed from the ship and towed back at relatively slow speeds (1-2 m/s). The relative distance between the transmitter and the receiver changed from 500 m to 7 km. In total 31 transmissions were carried out during the experiment, each containing 20 OFDM symbols in QPSK and 20 symbols in $16$ QAM, respectively. The transmitted OFDM signal adopts the same parameters shown in Table~\ref{tab.ofdm_para}. For data decoding, one transmission data file recorded during the turn of the source which has quite low SNR is excluded. In Fig.~\ref{mace10_chan_fig}, the estimated typical channel impulse response and the estimated Doppler speed corresponding to the 30 transmissions in MACE10 are presented.

\subsection{QPSK Decoding Results}
In total 600 OFDM symbols in QPSK modulation were collected during the sea experiment. Since all the original received data can be decoded successfully, additional noise in different level are added in order to show the differences between different algorithms. Denote the original noise variance in the received signal to be $\delta^2$, the decoding BER results corresponding to different additional noise levels are shown in Table~\ref{tab.mace10_dec_qpsk}. It can be observed that the proposed JCCD-VALSE algorithm always achieves the lowest BER in all additional noise levels.

\begin{table}[h]
\renewcommand\arraystretch{1.2}
  \centering
  \footnotesize
      \caption{MACE10 decoding BER: QPSK}\label{tab.mace10_dec_qpsk}
      \setlength{\tabcolsep}{5mm}{
    \begin{tabular}{|c|c|c|c|c|c|c|}
\hline
    Algorithm            & OMP    & AMP &  VALSE &  JCCD-VALSE & JCD-VALSE &  JCD-GAMP \\
 \Xhline{1.2pt}
    $0.8\delta^2$        & 0.0015	&  0.0021	& 0.0010	&  \textbf{0}	&  \textbf{0}	&  0.0001 \\  \hline
    $1.6\delta^2$        & 0.0085	&  0.0100	&  0.0078	&  \textbf{0.0056}	&  0.0058	&  0.0070 \\ \hline
    $2.4\delta^2$        & 0.0182	&  0.0194	&  0.0153	&  \textbf{0.0118}	&  \textbf{0.0118}	&  0.0125 \\ \hline
    $3.2\delta^2$        & 0.0338	&  0.0382	&  0.0281	&  \textbf{0.0204}	&  0.0214	&  0.0235 \\ \hline
    $4.0\delta^2$        & 0.0511	&  0.0568	&  0.0460	&  \textbf{0.0358}	&  0.0363	&  0.0397 \\ \hline
    $4.8\delta^2$        & 0.0662	&  0.0712	&  0.0616	&  \textbf{0.0531}	&  0.0535	&  0.0578 \\ \hline
    \end{tabular}
    }
    \normalsize
\end{table}

\subsection{$16$ QAM decoding results}
For 16 QAM modulation, 523 complete OFDM symbols were collected during the experiment. A direct decoding of the received signals with different algorithms leads to the results shown in Table~\ref{tab.mace10_dec_16QAM}, which clearly shows that the proposed JCCD-VALSE scheme achieves the best BER performance.

\begin{table}[h]
\renewcommand\arraystretch{1.2}
  \centering
  \footnotesize
      \caption{MACE10 decoding BER: 16 QAM}\label{tab.mace10_dec_16QAM}
      \setlength{\tabcolsep}{5mm}{
    \begin{tabular}{|c|c|c|c|c|c|c|}
\hline
    Algorithm            & OMP    & AMP &  VALSE &  JCCD-VALSE & JCD-VALSE &  JCD-GAMP \\
 \Xhline{1.2pt}
    BER        & 0.0152	& 0.0144	& 0.0130	& \textbf{0.0081}	& 0.0084	& 0.0131 \\  \hline
    \end{tabular}
    }
    \normalsize
\end{table}


\section{Conclusion}\label{Con}
This paper proposes the joint CFO, gridless channel estimation and data decoding algorithm named as JCCD-VALSE, which automatically estimates the number of paths, the nuisance parameters of the prior distribution and the noise variance. The algorithm is decomposed as three modules named the VALSE module, the minimum mean squared error (MMSE) module and the decoder module, and by iteratively exchange extrinsic information between modules, the channel estimation and data decoding performances gradually improve. Numerical experiments and real data are used to demonstrate the superior performance of the proposed approach.
\bibliographystyle{IEEEbib}
\bibliography{strings,refs}

\begin{thebibliography}{}
\bibitem{StPr09}
M. Stojanovic and J. Preisig, ``Underwater acoustic communication channels: Propagation models and statistical characterization,'' \emph{IEEE Commun. Mag.}, vol. 47, no. 1, pp. 84-89, Jan. 2009.
\bibitem{Sharif00JOE}
B. S. Sharif, J. Neasham, O. R. Hinton and A. E. Adams, ``A computationlly efficient Doppler compensation system for underwater acoustic communications,'' \emph{IEEE J. Ocean. Eng.}, vol. 25, no. 1, pp. 52-61, Jan. 2000.
\bibitem{Stoj06}
M. Stojanovic, ``Low complexity OFDM detector for underwater channels,'' \emph{in Proc. of MTS/IEEE OCEANS Conference}, Boston, MA, Sept. 18-21, 2006.
\bibitem{Zhou08JOE}
B. Li, S. Zhou, M. Stojanovic, L. Freitag and P. Willett, ``Multicarrier communication over underwater acoustic channels with nonuniform Doppler shifts,'' \emph{IEEE J. Ocean. Eng.}, vol. 33, no. 2, pp. 198-209, April 2009.
\bibitem{zhya20}
S. Zhao, S. Yan and J. Xi, ``Adaptive turbo equalization for differential OFDM systems in underwater acoustic communications,'' \emph{IEEE Trans. Veh. Technol.}, vol. 69, no. 11, pp. 13937-13941, 2020.
\bibitem{LiPr07}
W. Li and J. C. Preisig, ``Estimation of rapidly time-varying sparse channels,'' \emph{IEEE J. Ocean. Eng.}, vol. 32, no. 4, pp. 927-939, Oct. 2007.
\bibitem{bzpw10}
C. R. Berger, S. Zhou, J. C. Preisig, and P. Willett, ``Sparse channel estimation for multicarrier underwater acoustic communication: From subspace methods to compressed sensing,'' \emph{IEEE Trans. Signal Process.}, vol. 58, no. 3, pp. 1708-1721, 2010.
\bibitem{trgi07}
J. A. Tropp and A. C. Gilbert, ``Signal recovery from random measurements via orthogonal matching pursuit,'' \emph{IEEE Trans. Inform. Theory}, vol. 53, no. 12, pp. 4655-4666, 2007.
\bibitem{prdk15}
B. Peng, P. S. Rossi, H. Dong and K. Kansanen, ``Time-domain oversampled OFDM communication in doubly-selective underwater acoustic channels,'' \emph{IEEE Commun. Lett.}, vol. 19, no. 6, pp. 1081-1084, June 2015.
\bibitem{paup19}
E. Panayirci, M. T. Altabbaa, M. Uysal, and H. V. Poor, ``Sparse channel estimation for OFDM-based underwater acoustic systems in Rician fading with a new OMP-MAP algorithm,'' \emph{IEEE Trans. Signal Process.}, vol. 67, no. 6, pp. 1550-1565, 2019.
\bibitem{AMP}
D. L. Donoho, A. Maleki, and A. Montanari, ``Message passing algorithms for compressed sensing: I. Motivation and construction,'' \emph{in Proc. Inf. Theory Workshop}, Cairo, Egypt, Jan. 2010, pp. 1-5.
\bibitem{wlzd17}
C. Wei, H. Liu, Z. Zhang, J. Dang and L. Wu, ``Near-optimum sparse channel estimation based on least squares and approximate message passing,'' \emph{IEEE Wireless Commun. Lett.}, vol. 6, no. 6, pp. 754-757, Dec. 2017.
\bibitem{pasp10}
E. Panayirci, H. Senol and H. V. Poor, ``Joint channel Estimation, equalization, and data detection for OFDM systems in the presence of very high mobility,'' \emph{IEEE Trans. Signal Process.}, vol. 58, no. 8, pp. 4225-4238, Aug. 2010.
\bibitem{SchniterJAMP}
P. Schniter, ``A message-passing receiver for BICM-OFDM over unknown clustered-sparse channels,'' \emph{IEEE J. Sel. Topics Signal Process.}, vol. 5,
no. 8, pp. 1462-1474, Dec. 2011.
\bibitem{Schniter_AMP18}
P. Sun, Z. Wang and P. Schniter, ``Joint channel-estimation and equalization of single-carrier systems via bilinear AMP,'' \emph{IEEE Trans. Signal Process.}, vol. 66, no. 10, pp. 2772-2785, May 2018.
\bibitem{jswd20}
Z. Jiang, X. Shen, H. Wang and Z. Ding, ``Joint PSK data detection and channel estimation under frequency selective sparse multipath channels,'' \emph{IEEE Trans. Commun.}, vol. 68, no. 5, pp. 2726-2739, May 2020.
\bibitem{sshk08}
J. F. Sifferlen, H. C. Song, W. S. Hodgkiss, W. A. Kuperman and J. M. Stevenson, ``An iterative equalization and decoding approach for underwater acoustic communication,'' \emph{IEEE J. Ocean. Eng.}, vol. 33, no. 2, pp. 182-197, April 2008.
\bibitem{yazh16}
Z. Yang and Y. R. Zheng, ``Iterative channel estimation and Turbo equalization for multiple-input multiple-output underwater acoustic communications,'' \emph{IEEE J. Ocean. Eng.}, vol. 41, no. 1, pp. 232-242, Jan. 2016.
\bibitem{chwz17}
Z. Chen, J. Wang and Y. R. Zheng, ``Frequency-domain Turbo equalization with iterative channel estimation for MIMO underwater acoustic communications,'' \emph{IEEE J. Ocean. Eng.}, vol. 42, no. 3, pp. 711-721, July 2017.
\bibitem{qiqz21}
X. Qin, F. Qu and Y. R. Zheng, ``Bayesian iterative channel estimation and Turbo equalization for multiple-input multiple-output underwater acoustic communications,'' \emph{IEEE J. Ocean. Eng.}, vol. 46, no. 1, pp. 326-337, Jan. 2021.
\bibitem{armu18}
K. P. Arunkumar and C. R. Murthy, ``Iterative sparse channel estimation and data detection for underwater acoustic communications using partial interval demodulation,'' \emph{IEEE Trans. Signal Process.}, vol. 66, no. 19, pp. 5041-5055, Oct., 2018.
\bibitem{tuxs21}
X. Tu, X. Xu and A. Song, ``Frequency-domain decision feedback equalization for single-carrier transmissions in fast time-varying underwater acoustic channels,'' \emph{IEEE J. Ocean. Eng.}, vol. 46, no. 2, pp. 704-716, April 2021.
\bibitem{Rong20}
S. Wang, Z. He, K. Niu, P. Chen and Y. Rong, ``New results on joint channel and impulsive noise estimation and tracking in underwater acoustic OFDM systems,'' \emph{IEEE Trans. Wireless Commun.}, vol. 19, no. 4, pp. 2601-2612, April 2020.
\bibitem{ygdy21}
G. Yang, Q. Guo, H. Ding, Q. Yan and D. D. Huang, ``Joint message-passing-based bidirectional channel estimation and equalization with superimposed training for underwater acoustic communications,'' \emph{IEEE J. Ocean. Eng.}, early access, 2021.
\bibitem{MMis}
Y. Chi, L. L. Scharf, A. Pezeshki and R. Calderbank, ``Sensitivity of basis mismatch to compressed sensing,'' \emph{IEEE Trans. on Signal Process.}, vol. 59, pp. 2182 - 2195, 2011.
\bibitem{Supfast}
T. L. Hansen, B. H. Fleury and B. D. Rao, ``Superfast line spectral estimation,'' \emph{IEEE Trans. Signal Process.}, vol. 66, no. 10, pp. 2511-2526, 2018.
\bibitem{JointSupfast}
T. L. Hansen, P. B. J{\o}rgensen, M. A. Badiu and B. H. Fleury, ``An iterative receiver for OFDM with sparsity-based parametric channel estimation,'' \emph{IEEE Trans. Signal Process.}, vol. 66, no. 20, pp. 5454-5469, 2018.
\bibitem{Mengunified}
X. Meng, S. Wu and J. Zhu, ``A unified Bayesian inference framework for generalized linear models,'' \emph{IEEE Signal Process. Lett.}, vol. 25, no. 3, pp. 398-402, 2018.
\bibitem{Badiu}
M. A. Badiu, T. L. Hansen and B. H. Fleury, ``Variational Bayesian inference of line spectra,'' \emph{IEEE Trans. Signal Process.}, vol. 65, no. 9, pp. 2247-2261, 2017.
\bibitem{Direc}
K. V. Mardia and P. E. Jupp, \emph{Directional Statistics}. New York, NY, USA: Wiley, 2000.
\bibitem{Mackay1992}
D. J. C. MacKay, ``Bayesian interpolation,'' \emph{Neural Computation}, vol. 4, no. 3, pp. 415-447, 1992.
\bibitem{Berger1985}
J. O. Berger, \emph{Statistical Decision Theory and Bayesian Analysis}. Springer, second edition, 1985.
\bibitem{VMP}
J. Winn and C. M. Bishop, ``Variational message passing,'' \emph{J. Mach. Learn. Res.}, vol. 6, pp. 661-694, 2005.
\bibitem{Minka}
T. Minka, ``A family of algorithms for approximate Bayesian inference,'' Ph.D. dissertation, Department of Electrical Engineering and Computer Science, Mass. Inst. Technol., Cambridge, MA, USA, 2001.
\bibitem{ZZM18}
J. Zhu, Q. Zhang, and X. Meng, ``Grid-less variational Bayesian inference of line spectral from quantized samples,'' 2019, arXiv:1811.05680, to appear in \emph{China Communications}. [Online]. Available: https://arxiv.org/abs/1811.05680
\bibitem{wwzy12}
L. Wan, Z.-H. Wang, S. Zhou, T. C. Yang, and Z. Shi, ``Performance comparison of Doppler scale estimation methods for underwater acoustic OFDM,'' \emph{Journal of Electrical and Computer Engineering}, vol. 2012, Article ID 703243, 2012.
\bibitem{OMPBIC}
N. Eldarov, G. Tan, and T. Herfet, ``Delay-Doppler search for matching pursuit algorithms in time-variant channels,'' \emph{in Proc. IEEE Int. Symp.
Broadband Multimedia Syst. Broadcast.}, Jun. 2015, pp. 1-5.
\bibitem{EMAMPconf}
J. P. Vila and P. Schniter, ``Expectation-Maximization Bernoulli-Gaussian approximate message passing,'' \emph{Proc. Asilomar Conf. on Signals, Systems, and Computers}, Pacific Grove, CA, Nov. 2011.
\bibitem{EMAMP}
J. P. Vila and P. Schniter, ``Expectation-Maximization Gaussian-mixture approximate message passing,'' \emph{IEEE Trans. Signal Process.}, vol. 61, no. 19, pp. 4658-4672, Oct. 2013.
\bibitem{wjzm20}
L. Wan, H. Jia, F. Zhou, M. Muzzammil, T. Li and Y. Huang, ``Fine Doppler scale estimations for an underwater acoustic CP-OFDM system,'' \emph{Signal Process.}, vol. 170, May 2020.

\end{thebibliography}

\end{document}